# Antiferromagnetic spin-frustrated layers of corner-sharing Cu$_4$ tetrahedra on the kagome lattice in volcanic minerals Cu$_5$O$_2$(VO$_4$)$_2$(CuCl), NaCu$_5$O$_2$(SeO$_3$)$_2$Cl$_3$, and K$_2$Cu$_5$Cl$_8$(OH)$_4$·2H$_2$O

**L M Volkova and D V Marinin**

Institute of Chemistry, Far Eastern Branch, Russian Academy of Sciences 690022 Vladivostok, Russia

E-mail: volkova@ich.dvo.ru

arXiv:1804.10313v2

**Abstract**

The objective of the present work was to analyze the possibility of realization of quantum spin liquid in three volcanic minerals – averievite (Cu$_5$O$_2$(VO$_4$)$_2$(CuCl)), ilinskite (NaCu$_5$O$_2$(SeO$_3$)$_2$Cl$_3$), and avdononite (K$_2$Cu$_5$Cl$_8$(OH)$_4$·2H$_2$O) – from the crystal chemistry point of view. Based on the structural data, the sign and strength of magnetic interactions have been calculated and the geometric frustrations serving as the main reason of the existence of spin liquids have been investigated. According to our calculations, the magnetic structures of averievite and ilinskite are composed of antiferromagnetic (AFM) spin-frustrated layers of corner-sharing Cu$_4$ tetrahedra on the kagome lattice. However, the direction of nonshared corners of tetrahedra is different in them. The oxygen ions centering the OCu$_4$ tetrahedra in averievite and ilinskite provide the main contribution to the formation of AFM interactions along the tetrahedra edges. The local electric polarization in averievite and the possibility of spin configuration fluctuations due to vibrations of tetrahedra-centering oxygen ions have been discussed. The existence of structural phase transitions accompanied with magnetic transitions was assumed in ilinskite because of the effect of a lone electron pair by Se$^{4+}$ ions. As was demonstrated through comparison of averievite and avdoninite, at the removal of centering oxygen ions from tetrahedra, the magnetic structure of the pyrochlore layer present in averievite transformed into an openwork curled net with large cells woven from corner-sharing open AFM spin-frustrated tetrahedra ('butterflies') in avdoninite.

(Some figures may appear in colour only in the online journal)

# 1. Introduction

Search of potential materials characterized with realization of quantum spin liquid (QSL) [1-7] comprises an important task in the physics of condensed matter. At the present stage of studies, the frustration of magnetic interactions is considered as the main parameter, which must characterize possible candidates for QSL. Typical examples of geometrically frustrated magnets include antiferromagnets consisting of a two-dimensional (2D) triangular lattice and a three-dimensional (3D) pyrochlore lattice formed of corner-sharing tetrahedra. Compounds, in which magnetic ions form kagome lattices, are considered as the most promising as possible quantum spin liquids. At low temperatures, spin directions are usually ordered in dependence on the crystal lattice structure in substances with localized electrons. However, in lattices such as kagome, no ordering is observed at conventional (Heisenberg) interaction. On this lattice, spins form quantum spin liquid, i.e., they do not form an ordered structure even at very low temperatures. The available literature contains numerous theoretical evidences for QSLs in the system.

In spite of versatility of theoretical studies of QSL, there are a few known materials with such properties. A majority of low-dimensional systems studied up to present manifest a long-range order at low temperatures, even when interchain or interlayer exchange interactions are negligibly small in comparison with the main ones. Up to present, the existence of spin liquid is considered almost fully proven in herbertsmithite ($ZnCu_3(OH)_6Cl_2$) [8-11], whose magnetic system base comprises a kagome lattice composed of magnetic $Cu^{2+}$ ions. Besides, real tentative candidates for spin liquids include two more compounds: ruthenium chloride ($RuCl_3$) [12-16] (alpha modification) and complex calcium/chromium oxide ($Ca_{10}Cr_7O_{28}$) [17,18] having other structures of magnetic sublattices. The latter expands the prospects of search of new candidates of spin liquids among frustrated antiferromagnetics.

Versatility of structural motifs formed by magnetic atoms in Kamchatka volcanic minerals enables one to consider Kamchatka volcanos as a source of objects for search of extremely interesting magnetic materials for theoretical and practical applications [19-21]. Our preliminary studies [22, 23] indicated that frustration (disordering) of antiferromagnetic spin fragments was characteristic of Kamchatka-borne magnetic minerals, including low-dimensional quantum $S = 1/2$ spin systems with the $Cu^{2+}$ central ion containing antiferromagnetic pyrochlore layers, chains, or kagome lattices. The above structural fragments are responsible for frustration of exchange magnetic interactions because of geometric factors.

The present work is devoted to three minerals: averievite $Cu_5O_2(VO_4)_2(Cu^+Cl)$ [24], ilinskite $NaCu_5O_2(SeO_3)_2Cl_3$ [25], and avdoninite $K_2Cu_5Cl_8(OH)_4·2H_2O$ [26, 27]. The structure of the crystal sublattice of magnetic $Cu^{2+}$ ions in these minerals is composed of layers of corner-sharing $Cu_4$ tetrahedra located on the kagome lattice. In each mineral, these layers have specific features. In averievite and avdoninite, the crystal structure of layers is similar to that of the layers that can be cut from the 3D lattice of $Cu_2OSeO_3$ [28, 29] having a distorted pyrochlore structure [30, 31]. However, in averievite $Cu_4$ tetrahedra are centered by oxygen ions (just like in $Cu_2OSeO_3$), whereas in avdoninite they are void inside. The crystal structure of ilinskite has another specific feature – the direction of lone vertices of oxocentered tetrahedra in the layer is different from the pyrochlore structure.

To determine magnetic structures of these minerals, we calculated, based on the crystal structural data, characteristics (sign and strength) of magnetic interactions not only inside the low-dimensional fragments, but also between them at long distances and considered competition of these magnetic interactions on specific geometric configurations of the sublattice of magnetic $Cu^{2+}$ ions. According to our calculations, the structure of magnetic subsystem of the above minerals comprises antiferromagnetic (AFM) spin-frustrated layers of corner-sharing $Cu_4$ tetrahedra of three types located on the kagome lattice. The comparison of 2-D magnetic subsystems of these minerals was performed.

## 2. Method of calculation

The search of geometrically frustrated antiferromagnets containing layers of corner-sharing $Cu_4$ tetrahedra was performed in the Inorganic Crystal Structure Database (ICSD) among the minerals from volcanic fumaroles of the Tolbachik volcano (Kamchatka peninsula, Russia). To determine the characteristics of magnetic interactions (type of the magnetic moments ordering and strength of magnetic coupling) in minerals, we used the earlier developed method (named the 'crystal chemistry method') and the 'MagInter' program created on its basis [32-34]. Within the scopes of this method, three well-known concepts about the nature of magnetic interactions are used. First, it was the Kramers's idea [35], according to which the electrons of nonmagnetic ions play a considerable role in exchange couplings between magnetic ions separated by one or several diamagnetic groups. Second, we used the Goodenough–Kanamori–Anderson model [36-39], in which the crystal chemistry aspect clearly indicates to the dependence of the interaction strength and the type of orientation of spins of magnetic ions on the arrangement of intermediate anions. Third, we used the polar Shubin–Vonsovsky model [40]: considering magnetic interactions, we took into account not only anions, which are valence bound to the magnetic ions, but also all the intermediate negatively or positively ionized atoms, except cations of metals without unpaired electrons.

The method enables one to determine the sign (type) and strength of magnetic couplings on the basis of the structural data. According to this method, the coupling between magnetic ions $M_i$ and $M_j$ emerges at the moment of crossing the boundary between them by an intermediate ion ($A_n$) with the overlapping value of ~0.1 Å. The area of the limited space (local space) between the $M_i$ and $M_j$ ions along the bond line is defined as a cylinder, whose radius is equal to these ions radii. The strength of magnetic couplings and the type of magnetic moments ordering in insulators are determined mainly by the geometric position and the size of intermediate ions $A_n$ in the local space between two magnetic ions ($M_i$ and $M_j$). The positions of intermediate ions ($A_n$) in the local space are determined by the distance $h(A_n)$ from the center of the $A_n$ ion to the $M_i$–$M_j$ bond line and the degree of the ion displacement to one of the magnetic ions expressed as a ratio ($l_n'/l_n$) of the $l_n$ and $l_n'$ lengths ($l_n \leq l_n'$; $l_n' = d(M_i - M_j) - l_n$) produced by the $M_i$–$M_j$ bond line division by a perpendicular made from the ion center (figure 1).

The intermediate $A_n$ ions will tend to orient magnetic moments of $M_i$ and $M_j$ ions and make their contributions ($j_n$) into the emergence of AFM or ferromagnetic (FM) components of the magnetic interaction in dependence on the degree of overlapping of the local space between magnetic ions ($\Delta h(A_n)$), the asymmetry ($l_n'/l_n$) of position relatively to the middle of the $M_i$–$M_j$ bond line, and the distance between magnetic ions ($M_i$–$M_j$).

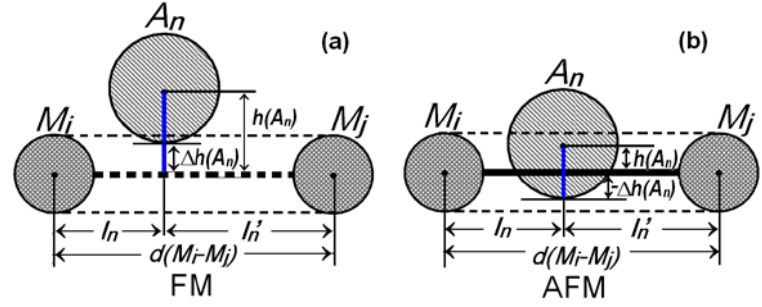

**Figure 1.** A schematic representation of the intermediate A*n* ion arrangement in the local space between magnetic ions M*i* and M*j* in cases where the A*n* ion initiates the emerging of the FM (**a**) and AFM (**b**) interactions. $h(An)$, $ln$, $ln'$, and $d(Mi–Mj)$ are the parameters determining the sign and strength of magnetic interactions ($Jn$).

Among the above parameters, only the degree of space overlapping between the magnetic ions $M_i$ and $M_j$ ($\Delta h(A_n) = h(A_n) - r_{A_n}$) equal to the difference between the distance $h(A_n)$ from the center of the $A_n$ ion to the $M_i$–$M_j$ bond line and the radius ($r_{A_n}$) of the $A_n$ ion determined the sign of magnetic interaction. If $\Delta h(A_n) < 0$, the $A_n$ ion overlaps (by $|\Delta h|$) the $M_i$–$M_j$ bond line and initiates the emerging contribution into the AFM-component of magnetic interaction. If $\Delta h(A_n) > 0$, there remains a gap (the gap width $\Delta h$) between the bond line and the $A_n$ ion, and this ion initiates a contribution to the FM-component of magnetic interaction. The sign and strength of the magnetic coupling ($J_{ij}$) are determined by the sum of the above contributions:

$$J_{ij} = \sum_n j_n$$

The $J_{ij}$ value is expressed in per angstrom units (Å$^{-1}$). If $J_{ij} < 0$, the type of $M_i$ and $M_j$ ions magnetic ordering is AFM and in opposite, if $J_{ij} > 0$, the ordering type is FM.

In spite of the rough character of the model, our method provides reasonable estimations not only on the spins' orientation, but also on the strength of the whole spectrum of magnetic couplings as inside the low-dimensional fragment as between the fragments. The method is sensitive to insignificant changes in the local space of magnetic ions and enables one to find intermediate ions localized in critical positions, deviations from which would result in the change of the magnetic coupling strength or spin reorientation (AFM–FM transition, for instance, under effect of temperature or external magnetic field).

Unlike *ab initio* methods, such as LDA or LDA-DMFT used in studies and simulation of the electronic structure of real strongly correlated systems, our method was created to search for magnetic compounds with the required magnetic structure based on the crystal structure data. As all the available methods, it has its own limitations and disadvantages, since, despite the determining role of structural factors in formation of the magnetic lattice, there exist other factors contributing to it. Besides, the results of our calculations strongly depend on the accuracy of determination of atomic coordinates and the compound composition. Slight deviations of the composition and structure of real crystals from the data on an ideal substance structure could result, in some cases, to substantial discrepancies with the experiment. However, the problems also emerge at using the most widely spread method (Monte-Carlo). For example, the authors of [41] showed that in the course of calculations of quantum kagome antiferromagnetics using this method it is principally impossible to eliminate the sign problem. The latter means that it is impossible to efficiently simulate such systems using conventional computers.

During calculations of the parameters of magnetic couplings and analysis of their competition on specific geometric configurations of the magnetic ions sublattice, we found three 2-D frustrated antiferromagnetics: averievite $Cu_5O_2(VO_4)_2(Cu^+Cl)$ [24], ilinskite $NaCu_5O_2(SeO_3)_2Cl_3$ [25], and avdoninite $K_2Cu_5Cl_8(OH)_4 \cdot 2H_2O$ [26]. The room-temperature structural data for averievite $Cu_5O_2(VO_4)_2(CuCl)$ (ICSD-85128), ilinskite $NaCu_5O_2(SeO_3)_2Cl_3$ (ICSD-188376), and avdoninite $K_2Cu_5Cl_8(OH)_4 \cdot 2H_2O$ (ICSD-55096) and ionic radii of Shannon [42] were used in the calculations. The radii of $^{IV}Cu^{2+}$, $^{V}Cu^{2+}$, $^{VI}O^{2-}$, $^{VI}Cl^-$, and $^{VI}Se^{4+}$ are equal to 0.57, 0.65, 1.40, 1.81, and 0.5 Å, respectively.

The minerals we investigate (averievite, ilinskite, and avdoninite) belong to the specific class of substances, whose magnetic structure and properties are to a great extent determined by two factors: the presence of Jahn–Teller $Cu^{2+}$ ions with an orbital degeneracy [43-45] and the geometric frustration of copper tetrahedra ($Cu_4$). Earlier [23], at studies of the magnetic structure of kamchatkite ($KCu_3OCl(SO_4)_2$), we demonstrated on the example of the $KCuF_3$ compound [46-50] that the intermediate F ions, the bond of copper with which is characterized with a Jahn–Teller elongation, did not contribute to the magnetic coupling. That is why at calculations of the parameters of magnetic couplings ($J$n) we did not take into account contributions from intermediate X ions ($j(X^{ax})$) having a direct elongated axial Cu–$X^{ax}$ bond with at least one of two $Cu^{2+}$ ions participating in the interaction.

To translate the $J$n value in per angstrom (Å$^{-1}$) into energy units more conventional for experimenters – millielectron-volt (meV) – one can use the average values of scaling factors ($K$n) ($K$ = 74) we calculated for the $J1$–$J4$ magnetic couplings in $Cu_3Mo_2O_9$ (figure 2, table 1). Using this scaling factor is possible according to two reasons. First, a single chain composed of corner-sharing oxocentered tetrahedra ($OCu_4$) in $Cu_3Mo_2O_9$ is a segment in averievite and ilinskite minerals, whereas the parameters of the $J1$–$J4$ magnetic couplings in this chain (at 1.5 K [51]) we calculated in Å$^{-1}$ are comparable with respective parameters in the minerals under examination. Second, the exchange interaction parameters ($J1$–$J4$) obtained from the experimental data were determined for $Cu_3Mo_2O_9$ (unit: meV) [52], which allowed calculation of the scaling factor.

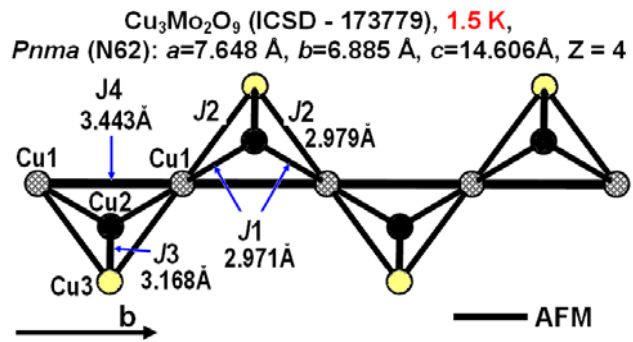

**Figure 2.** Single chains of corner-sharing tetrahedra ($Cu_4$) and the $J$n coupling in $Cu_3Mo_2O_9$.

**Table 1** An estimate of $J$n magnetic couplings in in single chains of corner-sharing $Cu_4$ tetrahedra in $Cu_3Mo_2O_9$ by crystal chemical method (unit: $Å^{-1}$) and experimental method [52] (unit: meV).

| | $Cu_3Mo_2O_9$ [51] (Data for ICSD – 173779 at 1.5 K) | | | |
|---|---|---|---|---|
| *Pnma* (N62): $a$ = 7.6476 Å, $b$ = 6.8855 Å, $c$ = 14.6058 Å $α$ =90°, $β$ = 90°, $γ$ = 90°, $Z$ = 4 | | | | |
| | $J1$ | $J2$ | $J3$ | $J4$ |
| Bond | Cu1-Cu2 | Cu1-Cu3 | Cu2-Cu3 | Cu1-Cu1 |
| d(Cu-Cu) (Å) | 2.971 | 2.979 | 3.168 | 3.443 |
| $J$n [a] ($Å^{-1}$) (AFM<0) | -0.0473 | -0.0430 | -0.0545 | -0.1161 |
| $J$n$^{exp}$ (meV)[b] (AFM>0) [52] | 3.06 | 3.06 | 5.7 | 6.5 |
| $K$n[c] | 64.69 | 71.16 | 104.59 | 55.99 |
| 74×$J$ ($Å^{-1}$) (AFM<0) | -3.50 | -3.18 | -4.03 | -8.59 |

[a]$J$n$^{str}$ – the magnetic couplings ($J$n < 0 - AFM, $J$n > 0 – FM) calculated on the basis of structural data (in $Å^{-1}$).
[b]$J$n – the exchange interaction parameters extracted from the experimental data (unit: meV) in $Cu_3Mo_2O_9$ [52].
[c]$K$n – scaling factors ($K$n = $J$n$^{exp}$ meV/$J$n$^{str}$ $Å^{-1}$) for translating the value $J$n$^{str}$ in per angstrom into meV; $K_{middle}$ = 74.

Table 2 shows the crystallographic characteristics and parameters of magnetic couplings ($J$n) calculated on the basis of the structural data in $Å^{-1}$ and respective distances between magnetic $Cu^{2+}$ ions in the materials under study.

## 3. Results and discussion

As was shown in [19, 53, 54], minerals of volcanic exhalations were formed under near-surface conditions with the participation of the fluid phase at high temperature and a pressure close to the atmospheric one. Many minerals of volcanic exhalations formed at the Great Tolbachik Fissure Eruption (GTFE), including averievite $Cu_5O_2(VO_4)_2(Cu^+Cl)$ and ilinskite $NaCu_5O_2(SeO_3)_2Cl_3$, are based on complexes composed of oxocentered $OCu_4$ tetrahedra. The presence of oxocentered $OCu_4$ tetrahedra in exhalation minerals and virtually instantaneous crystallization of the gas flow substance at its "impact" on the cold atmosphere allowed suggesting [55, 56] the mechanism of metal transfer by volcanic gases in the forms of such tetrahedra and their complexes and describing the process of formation of main mineral phases as an "assembly" of crystals from ready blocks. A majority of exhalation minerals are water-free compounds. However, many of them are unstable in air atmosphere. It is assumed that aqueous copper compounds are formed in the process of cooling of the erupted material at the expense of oxygen and humidity present in air. Among them, avdoninite $K_2Cu_5Cl_8(OH)_4·2H_2O$ was found [57]: $Cu_4$ tetrahedra also form its main structural units, but they are void.

### 3.1. Platform for frustrated magnetism in copper volcanic minerals

The common feature of the sublattice of magnetic ions in three minerals – averievite $Cu_5O_2(VO_4)_2(Cu^+Cl)$, ilinskite $NaCu_5O_2(SeO_3)_2Cl_3$, and avdoninite $K_2Cu_5Cl_8(OH)_4·2H_2O$ – is represented by layers of corner-sharing $Cu_4$ tetrahedra located on the kagome lattice. The triangular geometry of the sublattice of magnetic ions in these minerals can serve as the main reason of frustration of their magnetic subsystems, if magnetic couplings between the nearest-neighbors in

**Table 2.** Crystallographic characteristics and parameters of magnetic couplings ($J$n) calculated on the basis of structural data and respective distances between magnetic $Cu^{2+}$ ions in minerals: averievite $Cu_5O_2(VO_4)_2(Cu^+Cl)$, ilinskite $NaCu_5O_2(SeO_3)_2Cl_3$ and avdoninite $K_2Cu_5Cl_8(OH)_4 \cdot 2H_2O$.

| Crystallographic and magnetic parameters | $Cu_5O_2(VO_4)_2(Cu^+Cl)$ [24] (Data for ICSD - 85128) Space group $P3$ (N143) $a = b = 6.375$, $c = 8.399$ Å $\alpha = \beta = 90º, \gamma = 120º, Z=1$ Method[a] – XDS; R-value[b] = 0.052 | $NaCu_5O_2(SeO_3)_2Cl_3$ [25] (Data for ICSD - 188376) Space group $Pnma$ (N62) $a = 17.769, b = 6.448$, $c = 10.522$ Å $\alpha = \beta = \gamma = 90°, Z = 4$ Method[a] – XDS; R-value[b] = 0.044 | $K_2Cu_5Cl_8(OH)_4$ $2H_2O$ [26] (Data for ICSD - 55096) Space group $P2_1/c$ (N14) $a = 11.642, b = 6.564$, $c = 11.771$ Å $\alpha = \gamma = 90°, \beta = 91.09°, Z=4$ Method[a] – XDS; R-value[b] = 0.025 |
|---|---|---|---|
| | *Tetrahedron I* O2Cu3Cu2Cu2Cu2 | *Tetrahedron I* O1Cu2Cu3Cu4Cu4 | *Tetrahedron* Cu1Cu2Cu3Cu3 |
| Bond | Cu2-Cu3 | Cu2-Cu4 | Cu2-Cu3 |
| d(Cu-Cu) (Å) | 2.899 | 2.886 | 2.953 |
| $J$n[c] (Å$^{-1}$) | $J1 = -0.0822$ | $J1 = -0.0433$ | $J1 = -0.0182$ |
| $j(X)$[d] (Å$^{-1}$) | $j(O2)$: -0.0705 | $j(O1)$: -0.0459 | $j(O1)$: -0.0141 |
| ($\Delta h(X)$[e] Å, $l_n'/l_n$[f], CuXCu[g]) | (-0.295, 1.11, 105.27°) | (-0.189, 1.15, 99.25°) | (-0.061, 1.00, 95.60°) |
| $j(X)$[d] (Å$^{-1}$) | $j(O6)$: -0.0117 | $j(O3)$: 0.0026 | $j(O2)$: -0.0041 |
| ($\Delta h(X)$[e] Å, $l_n'/l_n$[f], CuXCu[g]) | (-0.049, 1.10, 93.97°) | (0.011, 1.09, 91.25°) | (-0.018, 1.02, 93.78°) |
| Bond | Cu2-Cu2 | Cu3-Cu4 | Cu2-Cu3 |
| d(Cu-Cu) (Å) | 3.144 | 2.961 | 3.276 |
| $J$n[c] (Å$^{-1}$) | $J2 = -0.0741$ | $J2 = -0.0661$ | $J2 = -0.0528$ |
| $j(X)$[d] (Å$^{-1}$) | $j(O2)$: -0.0741 | $j(O1)$: -0.0661 | $j(O1)$: -0.0528 |
| ($\Delta h(X)$[e] Å, $l_n'/l_n$[f], CuXCu[g]) | (-0.366, 1.00, 113.33°) | (-0.289, 1.08, 106.70°) | (-0.284, 1.02, 111.44°) |
| Bond | | Cu4-Cu4 | Cu1-Cu2 |
| d(Cu-Cu) (Å) | | 3.253 | 3.458 |
| $J$n[c] (Å$^{-1}$) | | $J6 = -0.1162$ | $J3 = -0.0665$ |
| $j(X)$[d] (Å$^{-1}$) | | $j(O1)$: -0.1162 | $j(O2)$: -0.0665 |
| ($\Delta h(X)$[e] Å, $l_n'/l_n$[f], CuXCu[g]) | | (-0.615, 1.00, 128.45°) | (-0.397, 1.05, 119.74°) |
| Bond | | Cu2-Cu3 | Cu1-Cu3 |
| d(Cu-Cu) (Å) | | 3.293 | 3.230 |
| $J$n[c] (Å$^{-1}$)] | | $J7 = -0.0724$ | $J4_1 = -0.0467$ |
| $j(X)$[d] (Å$^{-1}$) | | $j(O1)$: -0.0724 | $j(O2)$: -0.467 |
| ($\Delta h(X)$[e] Å, $l_n'/l_n$[f], CuXCu[g]) | | (-0.392, 1.05, 117.02°) | (-0.243, 1.04, 108.77°) |
| Bond | | | Cu3-Cu3 |
| d(Cu-Cu) (Å) | | | 3.302 |
| $J$n[c] (Å$^{-1}$) | | | $J5_1 = -0.0553$ |
| $j(X)$[d] (Å$^{-1}$) | | | $j(O1)$: -0.0553 |
| ($\Delta h(X)$[e] Å, $l_n'/l_n$[f], CuXCu[g]) | | | (-0.553, 1.02, 112.71°) |
| Bond | | | Cu1-Cu3 |
| d(Cu-Cu) (Å) | | | 3.605 |
| $J$n[c] (Å$^{-1}$) | | | $J6_1 = 0$ |
| | *Tetrahedron II* O3Cu1Cu2Cu2Cu2 | *Tetrahedron II* O2Cu1Cu3Cu4Cu4 | |
| Bond | Cu1-Cu2 | Cu1-Cu4 | |
| d(Cu-Cu) (Å) | 3.003 | 2.973 | |
| $J$n[c] (Å$^{-1}$) | $J3 = -0.0111$ | $J3 = -0.0377$ | |
| $j(X)$[d] (Å$^{-1}$) | $j(O3)$: -0.0282 | $j(O2)$: -0.0321 | |
| ($\Delta h(X)$[e] Å, $l_n'/l_n$[f], CuXCu[g]) | (-0.126, 1.15, 99.20°) | (-0.141, 1.07, 99.45°) | |
| $j(X)$[d] (Å$^{-1}$) | $j(O5)$: 0.0171 | $j(O6)$: -0.0056 | |
| ($\Delta h(X)$[e] Å, $l_n'/l_n$[f], CuXCu[g]) | (0.077, 1.00, 90.92°) | (-0.025, 1.03, 94.44°) | |

(*Continued.*)

**Table 2.** (*Continued.*)

| Crystallographic and magnetic parameters | Cu$_5$O$_2$(VO$_4$)$_2$(Cu$^+$Cl) [24] (Data for ICSD - 85128) Space group *P*3 (N143) $a = b = 6.375$, $c = 8.399$ Å $\alpha = \beta = 90°$, $\gamma = 120°$, Z=1 Method[a] – XDS; R-value[b] = 0.052 | NaCu$_5$O$_2$(SeO$_3$)$_2$Cl$_3$ [25] (Data for ICSD - 188376) Space group *Pnma* (N62) $a = 17.769$, $b = 6.448$, $c = 10.522$ Å $\alpha = \beta = \gamma = 90°$, Z = 4 Method[a] – XDS; R-value[b] = 0.044 | K$_2$Cu$_5$Cl$_8$(OH)$_4$ 2H$_2$O [26] (Data for ICSD - 55096) Space group *P*2$_1$/*c* (N14) $a = 11.642$, $b = 6.564$, $c = 11.771$ Å $\alpha = \gamma = 90°$, $\beta = 91.09°$, Z=4 Method[a] – XDS; R-value[b] = 0.025 |
|---|---|---|---|
| | *Tetrahedron II* O3Cu1Cu2Cu2Cu2 | *Tetrahedron II* O2Cu1Cu3Cu4Cu4 | |
| Bond | Cu2-Cu2 | Cu1-Cu3 | |
| d(Cu-Cu) (Å) | 3.233 | 3.168 | |
| $J_n$[c] (Å$^{-1}$) | $J4 = -0.0801$ | $J4 = -0.0656$ | |
| $j$(X)[d] (Å$^{-1}$) | $j$(O3): -0.0801 | $j$(O2): -0.0656 | |
| ($\Delta h$(X)[e] Å, $l_n'/l_n$[f], CuXCu[g]) | (-0.419, 1.00, 117.49°) | (-0.329, 1.00, 111.87°) | |
| Bond | | Cu4-Cu4 | |
| d(Cu-Cu) (Å) | | 3.195 | |
| $J_n$[c] (Å$^{-1}$) | | $J5 = -0.0433$ | |
| $j$(X)[d] (Å$^{-1}$) | | $j$(O2): -0.0433 | |
| ($\Delta h$(X)[e] Å, $l_n'/l_n$[f], CuXCu[g]) | | (-0.221, 1.00, 107.15°) | |
| Bond | | Cu3-Cu4 | |
| d(Cu-Cu) (Å) | | 3.341 | |
| $J_n$[c] (Å$^{-1}$) | | $J8 = -0.0709$ | |
| $j$(X)[d] (Å$^{-1}$) | | $j$(O2): -0.0709 | |
| ($\Delta h$(X)[e] Å, $l_n'/l_n$[f], CuXCu[g]) | | (-0.395, 1.05, 117.93°) | |
| *Kagomé lattice:* along sides of small triangles | | | |
| Bond | Cu2-Cu2 | Cu4-Cu4 | Cu3-Cu3 |
| d(Cu-Cu) Å | 6.375 | 6.448 | 6.564 |
| $J_n$[c] (Å$^{-1}$) | $J^{2+4}(J_{a,b}^{2-2}) = -0.0301$ | $J^{5+6}(J_b^{4-4}) = -0.0342$ | $J5_2 = J_b^{3-3-} = -0.0015$ |
| Bond | | Cu4-Cu4 | Cu3-Cu3 |
| d(Cu-Cu) | | 6.188 | 6.460 |
| $J_n$[c] (Å$^{-1}$) | | $J_{Cu4}^{2+8} = -0.0025$ | $J4_2 = -0.0313$ |
| Bond | | Cu3-Cu3 | Cu3-Cu3 |
| d(Cu-Cu) | | 6.270 | 7.210 |
| $J_n$[c] (Å$^{-1}$) | | $J_{Cu3}^{2+8} = 0.0011$ | $J6_2 = -0.0217 \leftrightarrow 0.0185$ |
| *Kagomé lattice:* large triangles | | | |
| Bond | Cu2-Cu2 | Cu4-Cu4 | Cu1-Cu3 |
| d(Cu-Cu) (Å) | 5.433 | 5.282 | 5.196 |
| $J_n$[c] (Å$^{-1}$) | $J5 = 0$ | $J9 = 0$ | $J7 = -0.0077$ |
| Bond | Cu2-Cu2 | Cu3-Cu4 | Cu3-Cu3 |
| d(Cu-Cu) (Å) | 5.609 | 5.420 | 5.997 |
| $J_n$[c] (Å$^{-1}$) | $J5' = 0$ | $J10 = 0$ | $J8 = -0.0053$ |
| Bond | | Cu3-Cu4 | Cu1-Cu3 |
| d(Cu-Cu) (Å) | | 5.669 | 6.287 |
| $J_n$[c] (Å$^{-1}$) | | $J11 = 0$ | $J9 = -0.0051$ |
| *Kagomé lattice:* diagonals | | | |
| Bond | Cu2-Cu2 | Cu4-Cu4 | Cu3-Cu3 |
| d(Cu-Cu) (Å) | 6.375 | 6.188 | 6.074 |
| $J_n$[c] (Å$^{-1}$) | $J6 = 0$ | $J12 = 0$ | $J10 = -0.0028$ |
| Bond | | Cu3-Cu3 | Cu1-Cu1 |
| d(Cu-Cu) (Å) | | 6.448 | 6.564 |

(*Continued.*)

| Crystallographic and magnetic parameters | $Cu_5O_2(VO_4)_2(Cu^+Cl)$ [24] | $NaCu_5O_2(SeO_3)_2Cl_3$ [25] | $K_2Cu_5Cl_8(OH)_4\ 2H_2O$ [26] |
|---|---|---|---|
| | (Data for ICSD - 85128) | (Data for ICSD - 188376) | (Data for ICSD - 55096) |
| | Space group $P3$ (N143) | Space group $Pnma$ (N62) | Space group $P2_1/c$ (N14) |
| | $a = b = 6.375$, | $a = 17.769$, $b = 6.448$, | $a = 11.642$, $b = 6.564$, |
| | $c = 8.399$ Å | $c = 10.522$ Å | $c = 11.771$ Å |
| | $\alpha = \beta = 90°$, $\gamma = 120°$, $Z = 1$ | $\alpha = \beta = \gamma = 90°$, $Z = 4$ | $\alpha = \gamma = 90°$, $\beta = 91.09°$, $Z=4$ |
| | Method[(a)] – XDS; | Method[(a)] – XDS; | Method[(a)] – XDS; |
| | R-value[(b)] = 0.052 | R-value[(b)] = 0.044 | R-value[(b)] = 0.025 |
| *Kagomé lattice:* diagonals | | | |
| $Jn^{(c)}$ (Å$^{-1}$) | | $J_b^{3-3} = 0$ | $J_b^{1-1} = 0.0026$ |
| Bond | | | Cu3-Cu3 |
| d(Cu-Cu) (Å) | | | 7.538 |
| $Jn^{(c)}$ (Å$^{-1}$) | | | $J11 = 0.0140$ |
| *Between adjacent tetrahedra in the layer* | | | |
| Bond | Cu2-Cu3 | Cu2-Cu3 | Cu1-Cu2 |
| d(Cu-Cu) | 5.356 | 4.777 | 5.427 |
| $Jn^{(c)}$ (Å$^{-1}$) | $J7 = -0.0404$ | $J13 = 0.0019$ | $J12 = -0.0052$ |
| Bond | Cu1-Cu2 | Cu1-Cu3 | Cu1-Cu23 |
| d(Cu-Cu) (Å) | 5.360 | 5.415 | 5.433 |
| $Jn^{(c)}$ (Å$^{-1}$) | $J8 = -0.0290$ | $J13' = 0.0011$ | $J13 = 0.0022$ |
| Bond | Cu2-Cu3 | Cu2-Cu4 | Cu2-Cu3 |
| d(Cu-Cu) (Å) | 5.417 | 5.379 | 4.979 |
| $Jn^{(c)}$ (Å$^{-1}$) | $J9 = -0.0520$ | $J15 = -0.0659$ | $J15 = 0.0003$ |
| Bond | Cu1-Cu2 | Cu1-Cu4 | Cu2-Cu3 |
| d(Cu-Cu) (Å) | 5.420 | 5.460 | 6.144 |
| $Jn^{(c)}$ (Å$^{-1}$) | $J10 = -0.0329$ | $J16 = -0.0703$ | $J16 = 0.0019$ |
| Bond | Cu1-Cu3 | Cu1-Cu2 | Cu2-Cu2 |
| d(Cu-Cu) | 5.901 | 5.306 | 6.915 |
| $Jn^{(c)}$ (Å$^{-1}$) | $J^{1+3} = -0.0334$ [-2.5] | $J^{4+7} = 0.0025$ | $J3_2 = -0.030 \leftrightarrow 0.022$ |
| *Between chains in the layer* | | | |
| Bond | Cu1-Cu3 | Cu1-Cu2 | Cu2-Cu2 |
| d(Cu-Cu) (Å) | 8.687 | 5.284 | 8.652 |
| $Jn^{(c)}$ (Å$^{-1}$) | $J11 = 0.0028$ | $J14 = -0.0041$ | $J14 = -0.0042$ |
| Bond | | Cu1-Cu4 | Cu2-Cu3 |
| d(Cu-Cu) (Å) | | 5.761 | 6.005 |
| $Jn^{(c)}$ (Å$^{-1}$) | | $J17 = -0.0002$ | $J17 = 0.0030$ |
| Bond | | Cu2-Cu4 | Cu2-Cu2 |
| d(Cu-Cu) (Å) | | 6.326 | 6.259 |
| $Jn^{(c)}$ (Å$^{-1}$) | | $J18 = -0.0013$ | $J18 = -0.0044$ |
| *Interlayers couplings* | | | |
| Bond | Cu1-Cu3 | Cu1-Cu1 | Cu2-Cu2 |
| d(Cu-Cu) (Å) | 5.281 | 4.779 | 7.499 |
| $Jn^{(c)}$ (Å$^{-1}$) | $J12 = 0.0033$ | $J19 = 0.0057$ | $J19 = -0.0042$ |
| Bond | Cu1-Cu2 | Cu2-Cu2 | Cu2-Cu2 |
| d(Cu-Cu) (Å) | 6.329 | 5.723 | 7.999 |
| $Jn^{(c)}$ (Å$^{-1}$) | $J13 = -0.0189$ | $J20 = -0.0020$ | $J20 = -0.0096$ |
| Bond | Cu2-Cu3 | Cu1-Cu4 | |
| d(Cu-Cu) (Å) | 6.401 | 6.212 | |
| $Jn^{(c)}$ (Å$^{-1}$) | $J14 = -0.0203$ | $J21 = -0.0024$ | |
| Bond | | Cu1-Cu3 | |
| | | 6.920 | |

(*Continued.*)
Table 2. (*Continued.*)

**Table 2.** (*Continued.*)

| Crystallographic and magnetic parameters | $Cu_5O_2(VO_4)_2(Cu^+Cl)$ [24] (Data for ICSD - 85128) Space group $P3$ (N143) $a = b = 6.375$, $c = 8.399$ Å $\alpha = \beta = 90°$, $\gamma = 120°$, Z=1 Method[a] – XDS; R-value[b] = 0.052 | $NaCu_5O_2(SeO_3)_2Cl_3$ [25] (Data for ICSD - 188376) Space group $Pnma$ (N62) $a = 17.769$, $b = 6.448$, $c = 10.522$ Å $\alpha = \beta = \gamma = 90°$, Z = 4 Method[a] – XDS; R-value[b] = 0.044 | $K_2Cu_5Cl_8(OH)_4\ 2H_2O$ [26] (Data for ICSD - 55096) Space group $P2_1/c$ (N14) $a = 11.642$, $b = 6.564$, $c = 11.771$ Å $\alpha = \gamma = 90°$, $\beta = 91.09°$, Z=4 Method[a] – XDS; R-value[b] = 0.025 |
|---|---|---|---|
| d(Cu-Cu) (Å) | | 6.920 | |
| $Jn$[c] (Å$^{-1}$) | | $J22 = -0.0031$ | |
| Bond | | Cu2-Cu4 | |
| d(Cu-Cu) (Å) | | 7.242 | |
| $Jn$[c] (Å$^{-1}$) | | $J23 = -0.0108$ | |
| Bond | | Cu1-Cu2 | |
| d(Cu-Cu) (Å) | | 7.408 | |
| $Jn$[c] (Å$^{-1}$) | | $J24 = -0.0049$ | |

[a]XDS – X-ray diffraction from single crystal.
[b]The refinement converged to the residual factor (*R*) values.
[c]$Jn$ in Å$^{-1}$ ($Jn$ (meV) = $Jn$ (Å$^{-1}$)×K, where scaling factors $K_{middle}$ = 74) – the magnetic couplings ($Jn$<0 - AFM, $Jn$>0 – FM).
[d]$j(X)$ – contributions of the intermediate X ion into the AFM ($j(X) <0$) and FM ($j(X)>0$) components of the $Jn$ coupling
[e]$\Delta h(X)$ – the degree of overlapping of the local space between magnetic ions by the intermediate ion X.
[f]$l_n'/l_n$ – the asymmetry of position of the intermediate X ion relatively to the middle of the $Cu_i$–$Cu_j$ bond line.
[g]$Cu_iXCu_j$ – bonding angle.

triangles are of the antiferromagnetic nature compatible in strength (figure 3). In this case, the simultaneous energy minimization for all pairwise interactions in triangles forming both tetrahedra and kagome lattice is impossible. This frustration causes the spins to constantly fluctuate between different arrangements. Upon fluctuations, the resulting liquid-like state strongly attenuates the effective exchange interaction, which complicates (in some cases, makes impossible) the formation of a long-range magnetic order until low temperatures and, finally, can promote realization of the spin–liquid state. The dependence of the nearest neighbor interactions on the M–X–M bonding angle is proven and generally accepted. In case of oxocentered $OCu_4$ tetrahedra, the Cu–O–Cu angle value exceeds, as a rule, 90° (in a regular oxocentered tetrahedra it is equal to 109.5°), which allows immediate assumption on the antiferromagnetic character of interaction along edges of such tetrahedra.

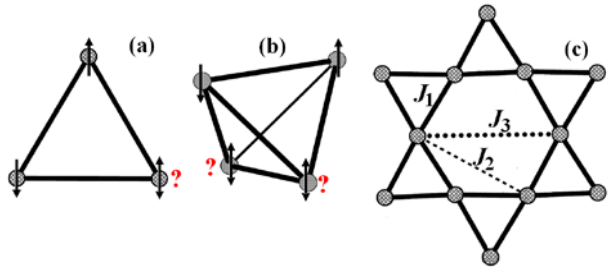

**Figure 3.** Frustrated "plaquettes": the triangle (a), the tetrahedron (b). Sketch of the kagome lattice and interactions between first ($J_1$), second ($J_2$), and third ($J_3$) nearest-neighbors.

The frustrated Heisenberg model of the pyrochlore lattice composed of corner-sharing tetrahedra comprises a 3D archetype of a frustrated lattice. It is well-known that the nearest-neighbor model is, apparently, extremely frustrated, just like the related 2D kagome model. However, because of the 3D character, since it cannot be numerically simulated, there are just a few results of such calculations [3]. In opposite, studies of the frustration of magnetic interactions on the kagome lattice have been described in numerous works, and very interesting results have been obtained. To study disordered quantum ground states of spin systems, the spin-1/2 Heisenberg model on the kagome lattice [3] was used the most actively. First of all, the minimal model with a single AFM exchange ($J_1$) on the nearest-neighbor bonds, then the effects of second- and third-neighbor exchange ($J_2$ and $J_3$) were studied (figure 3(c)), as such a behavior of frustrated systems strongly depends on interaction not only with the nearest, but also with farther neighbors. The available literature data are insufficient to create a general picture of this dependence. Let us describe just a few facts about the effect of interactions with farther neighbors on the frustration degree, which are of special interest for the present study. As was shown by studies of the phase diagram of the $J_1$-$J_2$ Heisenberg model on the kagome lattice by Kolley et al. [58], the magnetic order in the range -0.1≤$J_2$≤0.2 was absent in a narrow interval around $J_2$ ~ 0, which was compatible with the spin-liquid behavior. Here, it was accepted that AFM $J_1$ = 1, whereas for $J_2$ both antiferromagnetic and ferromagnetic states were considered. As was shown in [59], a substantial spin-liquid phase was centered near $J_2$ = 0.05–0.15, while in [60] the limits of the existence of spin liquid were expanded to $J_2/J_1$≤0.3. In the case of kagome, the narrowest range of stability for the existence of the gapless spin-liquid ground state (-0.03 ≤$J_2/J_1$≤ 0.045) is presented in [61]. The simulations [62-65] including the third-neighbor exchange (here, $J_3$ is defined as the exchange between sites on opposite corners of a hexagon of the lattice) (Fig. 3c) revealed a chiral QSL state in the range 0.3<$J_2$≈$J_3$<0.7.

It appears difficult to determine the hierarchy of magnetic interactions in the minerals under examination, in which individual magnetic fragments in the form of $Cu_4$ tetrahedra and a kagome lattice were linked into layers. In addition to the nearest interactions in layers, even very weak long-range interactions (also between layers) could have a strong effect on the magnetic state of the frustrated quantum magnet at low temperatures [66, 67].

Hereafter, we will discuss the parameters of magnetic interactions and their competition on specific geometric fragments and attempt to estimate the possibility of realization of the spin liquid state type in three quasi-two-dimensional systems: averievite $Cu_5O_2(VO_4)_2(Cu^+Cl)$, ilinskite $NaCu_5O_2(SeO_3)_2Cl_3$, and avdoninite $K_2Cu_5Cl_8(OH)_4·2H_2O$ from the perspective of crystal chemistry.

### 3.2. Averievite $Cu_5O_2(VO_4)_2(Cu^+Cl)$

Averievite $Cu_5O_2(VO_4)_2(CuCl)$ [24] crystallizes in the noncentrosymmetric trigonal $P$3 system. Magnetic $Cu^{2+}$ ions occupy 3 crystallographically independent sites (Cu1, Cu2, and Cu3) and have characteristic distortions of $Cu^{2+}$ coordination polyhedra ($Cu1O_5$ – a trigonal bipyramid, where d(Cu-O) = 1.806 – 2.109 Å; $Cu2O_4$ – a distorted square, where d(Cu2-O) = 1.881 – 2.104 Å; $Cu3O_5$ – a trigonal bipyramid, where d(Cu-O) = 1.765 – 2.032 Å) due to the Jahn–Teller effect strengthened by geometric hindrances related to the packing features. The base of the crystal structure of averievite is formed by two types of oxocentered corner-sharing $OCu_4$ tetrahedra (tetrahedron I – *O2Cu3Cu2Cu2Cu2* and tetrahedron II – *O3Cu1Cu2Cu2Cu2*) (figures 4(a) and (d)), linked in $[O_2Cu_5]^{6+}$ into layers (figure 5). The oxocentered layers are located exactly one above another.

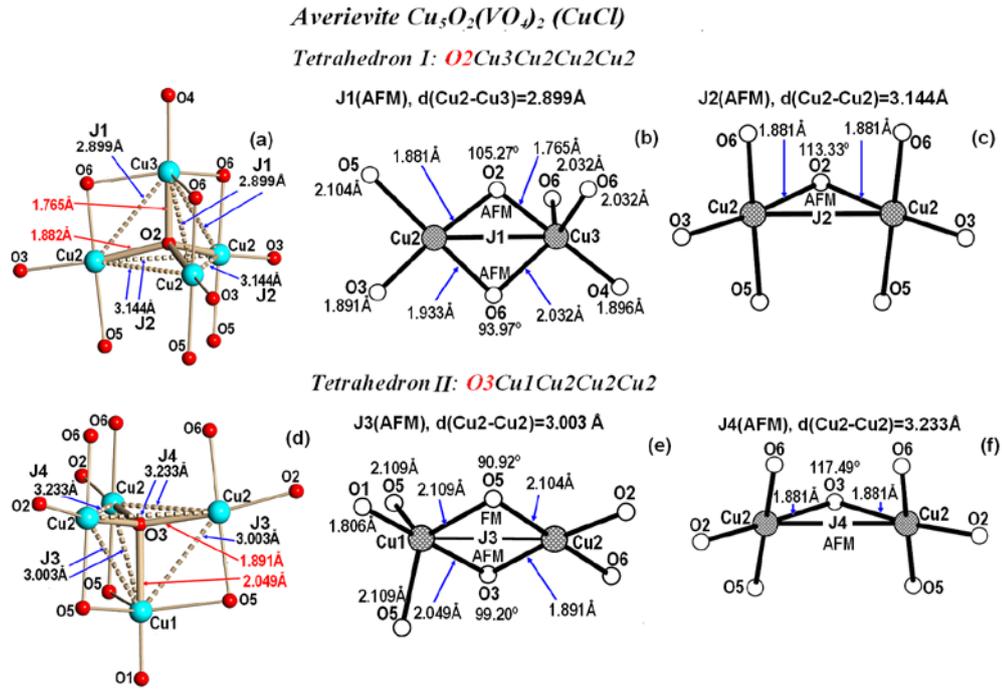

**Figure 4.** Linking the CuOm coordination polyhedra into tetramers *I* (a) and *II* (d) in averievite $Cu_5O_2(VO_4)_2(CuCl)$. The arrangement of intermediate ions in the local space of AFM *J*1 (b), *J*2 (c) in tetrahedron *I* and *J*3 (e) and *J*4 (f) in tetrahedron II.

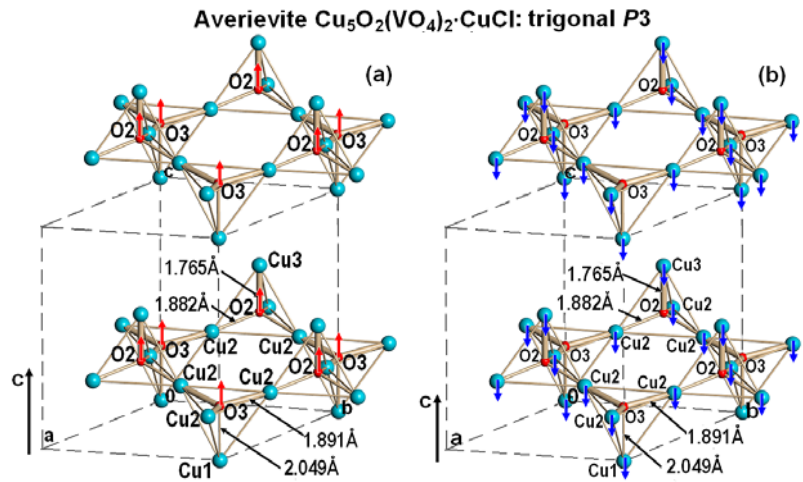

**Figure 5.** The $[O2Cu5]^{6+}$ layers in the structures of averievite. Electric polarization of $OCu_4$ tetrahedra. Separation of centers of gravity of positive and negative charges: (a) displacement of O2 and O3 oxygen ions along the 001 direction; (b) displacement of the sublattice of Cu1 and Cu3 copper ions along the 00-1 direction. In this and other figures, the thick and thin lines refer to short and long Cu-O bonds, respectively.

The crystal sublattice of magnetic ions in averievite comprises unbound to each other layers composed of corner-sharing $Cu_4$ tetrahedra, in which lone Cu1 and Cu3 vertices of adjacent tetrahedra I and II, respectively, are directed to opposite sides relatively to the plane of the layer formed by Cu2 ions (figures 6(a) and (b)). These layers can be considered as geometric isomers of layers "cut" from the 3-D pyrochlore lattice of $Cu_2OSeO_3$ [28, 29]. If one divides averievite layers into simple chains, in which lone vertices of adjacent tetrahedra are directed to opposite sides, then $Cu_3Mo_2O_9$ chains will serve as geometric isomers [51, 52] (figure 2).

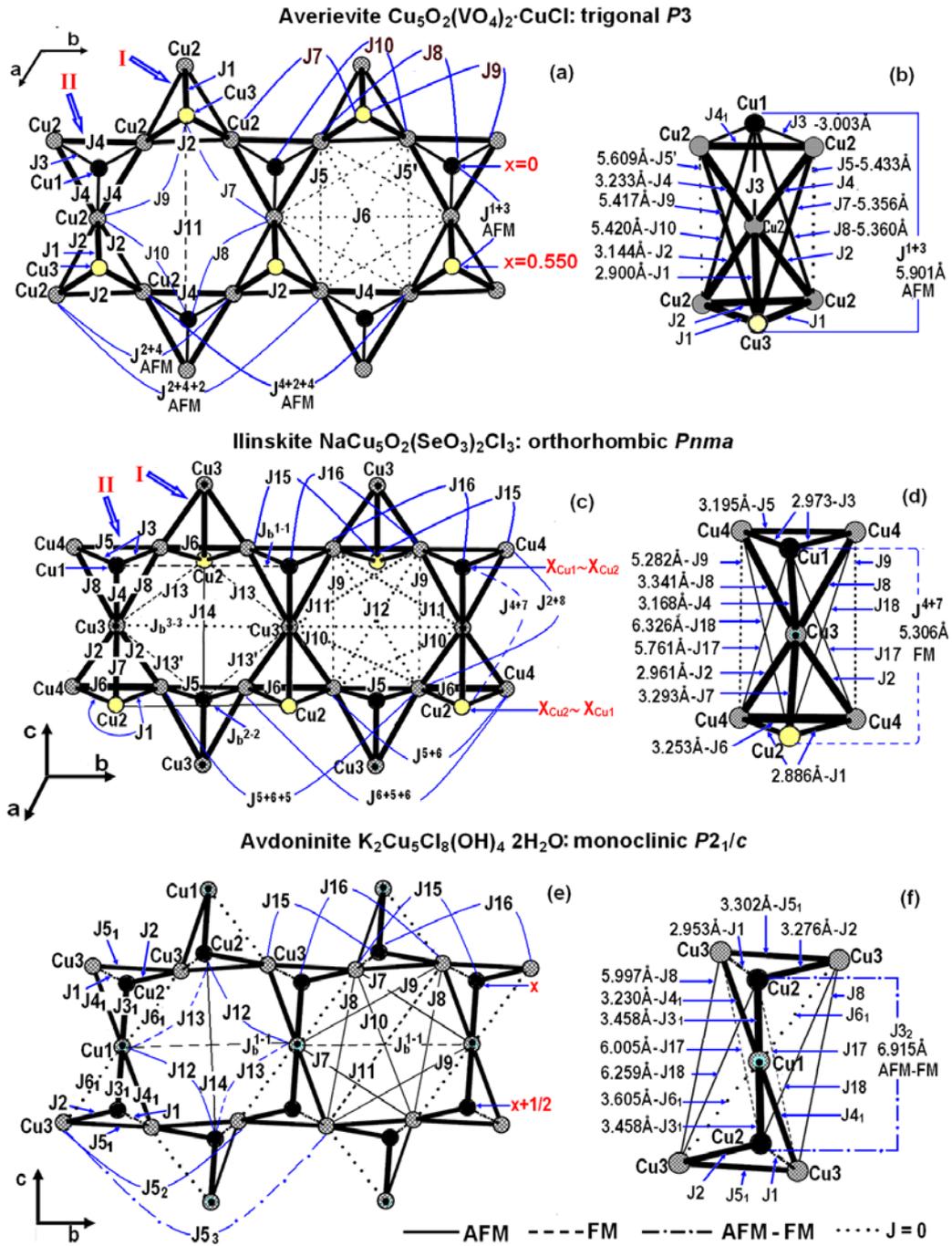

**Figure 6.** Intralayer $J_n$ couplings in averievite $Cu_5O_2(VO_4)_2(Cu^+Cl)$ (a, b), ilinskite $NaCu_5O_2(SeO_3)_2Cl_3$ (c, d), and avdoninite $K_2Cu_5Cl_8(OH)_4 \cdot 2H_2O$ (e, f). In this and other figures, the *thickness of lines* shows the strength of $J_n$ couplings. AFM and FM couplings are indicated by *solid* and *dashed lines*, respectively. The possible FM→AFM transitions are shown by the *stroke in dashed lines*

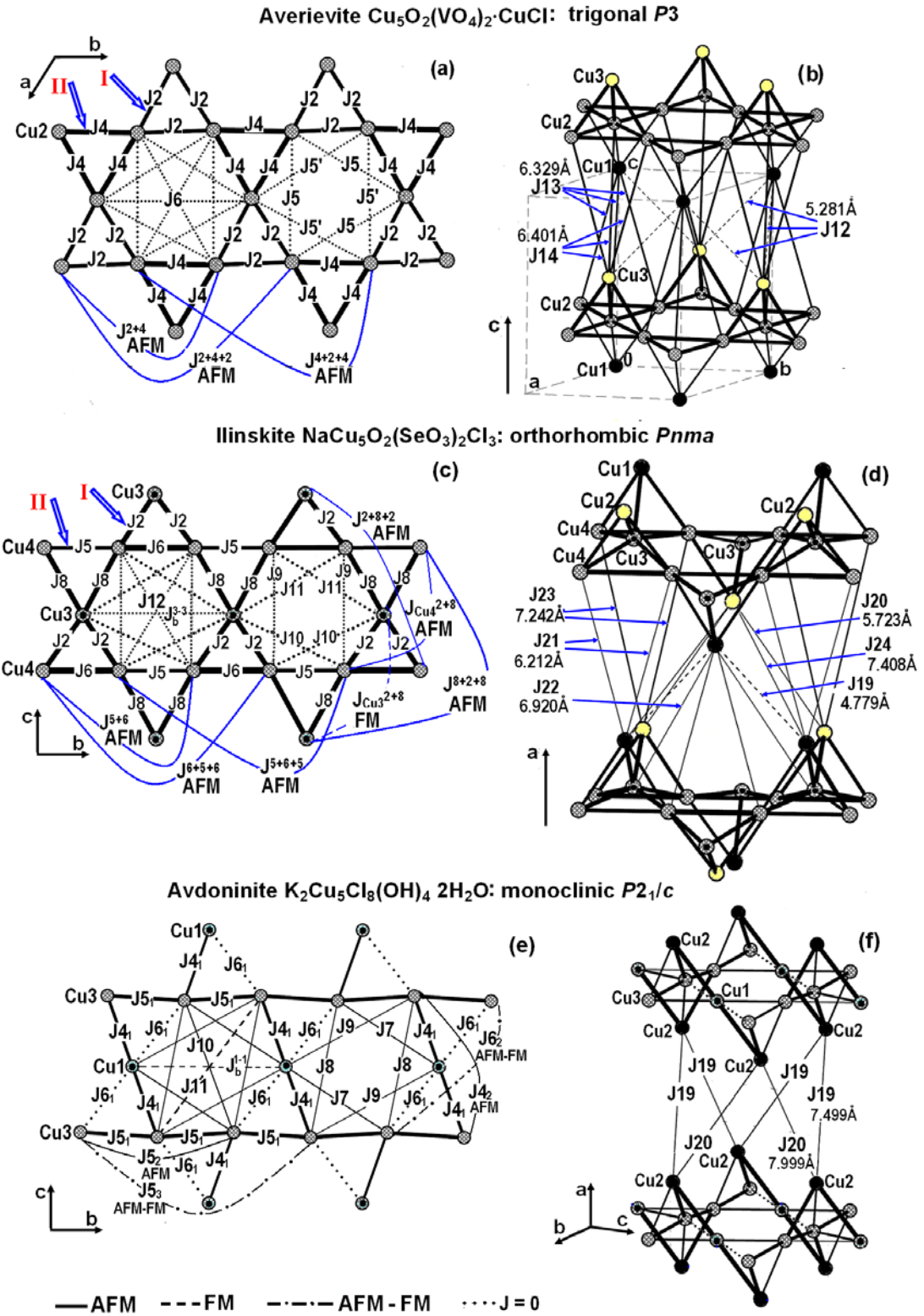

**Figure 7.** *Jn* couplings in the kagome lattice and between layers in averievite $Cu_5O_2(VO_4)_2(Cu^+Cl)$ (a, b), ilinskite $NaCu_5O_2(SeO_3)_2Cl_3$ (c, d), and avdoninite $K_2Cu_5Cl_8(OH)_4 \cdot 2H_2O$ (e, f).

Finally, the sublattice of $Cu^{2+}$ ions in averievite can be considered as a kagome lattice formed by the Cu2 ions, in which small triangles are centered by the Cu1 and Cu3 copper ions from opposite sides. As a result, a half of triangles are centered by the Cu1 ions from above, while another half, in opposite, are centered by the Cu3 ions from below (figure 6(a)).

Let us consider what characteristics of magnetic couplings can be present in averievite, if their formation were caused exclusively by the crystal structure. According to our calculations (table 2, figures 6(a) and (b), 7(a)), the nearest $J2$ ($J2$ = -0.0741 Å$^{-1}$, d(Cu2-Cu2) = 3.144 Å) and $J4$ ($J4$ = -0.0801 Å$^{-1}$, d(Cu2-Cu2) = 3.233 Å) couplings between Cu2 ions in the kagome lattice are strong AFM couplings and compete in $J2$–$J2$–$J2$ and $J4$–$J4$–$J4$ triangles forming bases of the tetrahedra *I* and *II*, respectively. These $J2$ and $J4$ couplings are formed under effect of the O2 ion centering the tetrahedron *I* and the O3 ion centering the tetrahedron *II*, respectively (figures 4cf). An additional competition in the kagome lattice occurs between the nearest AFM $J2$ ($J^{2+4}/J2$ = 0.41) and $J4$ ($J^{2+4}/J4$ = 0.38) couplings with the next-nearest AFM $J^{2+4}$ ($J^{2+4}$ = -0.301, d(Cu2-Cu2) = 6.375 Å = *a*) couplings in the chain along the sides of small triangles.

Just small triangles are frustrated in the kagome plane. There is no frustration in the kagome lattice honeycombs, since there are no couplings at long distances along the sides of large triangles ($J5$ (d(Cu2-Cu2) = 5.433 Å) and $J5'$(d(Cu2-Cu2) = 5.609 Å)) and along diagonals ($J6$ (d(Cu2-Cu2) = 6.375Å)) (see figures 6(a), 7(a), table 2).

If one assumes that the Cu1 and Cu3 copper ions building up the kagome lattice until the pyrochlore layer are absent, the presence of strong AFM $J2$ and $J4$ nearest-neighbor couplings and equal to 0 values of the $J5$ and $J5'$ second- and $J6$ third-neighbor couplings indicate to the possibility of the existence of the spin-liquid phase (-0.03 ≤$J_2/J_1$≤ 0.045), according to [61].

However, in our case, we cannot limit ourselves with just the above data, as the kagome lattice in averievite comprises just one of the components of the pyrochlore layer. The nearest couplings of the Cu3 and Cu1 ions with the Cu2 ones from the kagome lattice increasing small triangles until tetrahedra I and II, respectively, participate to full extent in formation of the magnetic state of the pyrochlore layer. Three $J1$ couplings ($J1$ = -0.0822 Å$^{-1}$, d(Cu2-Cu3) = 2.899 Å) along the edges of the tetrahedron I between the Cu3 and Cu2 ions (figures. 4(a) and (b), 6(a)) are the strongest AFM couplings in the averievite structure ($J1/J2$ = 1.11). The main contribution to formation of the AFM character of these couplings is provided by the O2 oxygen ion centering this tetrahedron. In addition, a small contribution to the AFM component of the $J1$ coupling is provided by the O6 ion entering the local space of this interaction, although it is located outside the tetrahedron I. However, this contribution could disappear completely at displacement of the O6 ion by just 0.05 Å from the Cu2–Cu3 bond line (critical position 'b' of the intermediate ion ($h(A) \approx r_A$ ($\Delta h(A) \approx 0$)) [34]).

Three AFM $J3$ ($J3$ = -0.0111 Å$^{-1}$, d(Cu1-Cu2) = 3.003 Å) couplings along the edges of the tetrahedron *II* between the Cu1 and Cu2 ions (figures 4(d) and (e), 6(a)) are the weakest ($J3/J4$ = 0.14) and the most unstable among the nearest couplings. The reason here consists in the fact that formation of the $J3$ coupling is contributed by two intermediate O3 and O5 couplings. These contributions are just slightly different in the value, but opposite in sign and, therefore, attenuate each other to some degree (critical positions 'd' [34]). In this case, a slight displacement of even one of the intermediate ions could result in complete disappearance of the magnetic coupling or the AF–FM transition. Thus, the competition between AFM couplings in the kagome lattice itself is supplemented by a strong competition between AFM $J1$ and $J2$ couplings in the tetrahedron I, whereas the competition between AFM *J*3 and *J*4 couplings in the tetrahedron II is very weak. However, this competition could increase, if one slightly shifts the O3 ion from the center of the tetrahedron II from the kagome plane (see below).

Four comparatively strong AFM *J*7 (*J*7 = -0.040 Å$^{-1}$, d(Cu2-Cu3) = 5.356 Å), *J*8 (*J*8 = -0.029 Å$^{-1}$, d(Cu1-Cu2) = 5.360 Å), *J*9 (*J*9 = -0.052 Å$^{-1}$, d(Cu2-Cu3) = 5.417 Å) and *J*10 (*J*10 = -0.033 Å$^{-1}$, d(Cu1-Cu2) = 5.420 Å) couplings exist between copper ions from adjacent tetrahedra (table 2, figures 6(a) and (b)). The main contribution in formation of the *J*7 and *J*9 couplings is provided by the intermediate O6 ions, while the *J*8 and *J*10 couplings are formed under effect of the O5 ions. Both of these oxygen ions are located outside the tetrahedra. One more AFM *J*$^{1+3}$ (*J*$^{1+3}$ = -0.0334 Å$^{-1}$, d(Cu1-Cu3) = 5.901 Å) coupling between the Cu1 and Cu3 ions is formed under effect of the intermediate Cu2 ion located virtually in the middle of the bond between these ions. The above AFM couplings between adjacent tetrahedra form six AFM triangles [*J*2*J*8*J*10 (*J*8/*J*2 = 0.39, *J*10/*J*2 = 0.44); *J*4*J*7*J*9 (*J*7/*J*4 = 0.50, *J*9/*J*4 = 0.65) *J*3*J*$^{1+3}$*J*7 (*J*$^{1+3}$/*J*3 = 3.0, *J*7/*J*3 = 3.6); *J*3*J*$^{1+3}$*J*9 (*J*$^{1+3}$/*J*3 = 3.0, *J*9/*J*1 = 4.7), *J*1*J*$^{1+3}$*J*10 (*J*$^{1+3}$/*J*1 = 0.41, *J*10/*J*1 = 0.39); *J*1*J*$^{1+3}$*J*8 (*J*$^{1+3}$/*J*1 = 0.41, *J*8/*J*1 = 0.35] and, as a result, create an additional competition to the layers.

Between the layers, one observes the AFM *J*13 (*J*13/*J*4 = 0.24, d(Cu1-Cu2) = 6.329 Å) and AFM *J*14 (*J*14/*J*2 = 0.27, d(Cu3-Cu2) = 6.401 Å) couplings, which could insignificantly compete with the AFM *J*4 and *J*2 couplings from the kagome lattice, and the very weak FM *J*12 (*J*12 = 0.0033 Å$^{-1}$, d(Cu1-Cu3) = 5.281 Å) couplings (table 2, figure 7(b)).

Thus, from the point of crystal chemistry, the magnetic structure of averievite is composed of AFM spin-frustrated layers of corner-sharing Cu$_4$ tetrahedra on the kagome lattice. Moreover, if one takes into account the known models of existence of the quantum spin liquid on the kagome lattice, then the calculations of the parameters of magnetic couplings we performed enable one to suggest that liquid existence on an individual fragment in averievite, namely, on the kagome lattice. It is worth emphasizing that the O2 ions centering the tetrahedron I (O2Cu3Cu2Cu2Cu2) (figures 4(a), (b) and (c)) and the O3 ions centering the tetrahedron II (O3Cu1Cu2Cu2Cu2) (figures 4(d)-(f)) provide the main contribution to antiferromagnetic components of spin interactions in OCu$_4$ tetrahedra. Let us further consider what effects one can expect even at slight changes in positions of oxygen ions centering the Cu$_4$ tetrahedra. The shift of the O2 and O3 ions in the same direction could result in a local electric polarization of [O$_2$Cu$_5$]$^{6+}$ layers in the structures of averievite. Vibrations of the O2 and O3 ions inside the tetrahedra could result in fluctuations of spin configurations, which will be considered in more detail below.

*3.2.1. Local electric polarization in averievite.* A local electric polarization exists in averievite along the *c* axis. Separation of the gravity centers of positive and negative charges is expressed in the shift of the oxygen O2 (by 0.09 Å) and O3 (by 0.13 Å) ions centering the tetrahedra I (O2Cu3Cu2Cu2Cu2) and II (O3Cu1Cu2Cu2Cu2), respectively, along the 001 direction from the tetrahedra centers (figures 5(a) and 8(a)). As a result, the bond lengths in the tetrahedra I and II are shortened (O2-Cu3 = 1.77 Å and O3-Cu2 = 1.89 Å) along the 001 direction and elongated (O2-Cu2 = 1.88 Å and O3-Cu1 = 2.05 Å) along the opposite 00-1 direction. The latter can be also considered as a shift of the copper ions sublattice (Cu1 на 0.144 Å, Cu2 на 0.084 Å and Cu3 на 0.095 Å) along the 00-1 direction (figure 5b). The electric polarization can be eliminated by two methods: first, shifting the O2 (from the initial z(O2) = 0.3390 to z(O2) = 0.3278) and O3 (from the initial z(O3) = 0.2440 to z(O3) = 0.2283) ions in the 00-1 direction; second, shifting the Cu1 (from the initial z(Cu1) = 0 to z(Cu1) = 0.0172), Cu2 (from the initial z(Cu2) = 0.280 to z(Cu2) = 0.290), and Cu3 (from the initial z(Cu3) = 0.5492 to z(Cu3) = 0.5606) ions in the opposite 001 direction. As a result, the lengths of four O2-4Cu bonds in the tetrahedron *I* and O3-4Cu ones in the tetrahedron *II* will become equal to 1.86 and 1.92 Å, respectively.

We have calculated the parameters of *J*1*-*J*4* magnetic couplings for the hypothetical location of the centering O2 and O3 ions in the centers of the tetrahedra *I* and *II*, respectively, and compared these data to those for the initial polarized structure. It turns out that, in all cases except one, the parameters of magnetic couplings in the unpolarized structure differ insignificantly from respective parameters of the initial polarized structure. The most significant changes occur upon shifting of the O3 ion (by 0.036 Å) to the center of the tetrahedron *II* from the kagome plane. If the O3 ions were located in the centers of the tetrahedra *II*, the strength of the AFM *J*3 couplings of the Cu1 ions with the Cu2 ones forming the kagome lattice would increase 2.6-fold, and the ratio (*J*3/*J*4 = 0.41) of the strengths of the AFM *J*3 and *J*4 couplings would exceed the critical value α = 0.25. Here, the competition between the AFM *J*3 and *J*4 couplings increased substantially and attained a frustration. However, the shift of the O3 ion from the center to the kagome plane results in the decrease (1.6-fold) of its contribution *j*(O3) to the AFM component of the *J*3 coupling and, in contrast, to an insignificant decrease (1.13-fold) of the *J*4 coupling, which results in the decrease of the *J*3/*J*4 ratio down to 0.14. Therefore, in case of averievite, the presence of polarization results in a significant weakening of competition between the *J*3 and *J*4 couplings in the tetrahedron *II* (O3Cu1Cu2Cu2Cu2) in comparison with the hypothetical centrosymemtric variant (figures 6(a) and (b)). Possibly, in this quasi-two-dimensional spin-frustrated system, the elimination of the symmetry center and the emergence of a local electric polarization could occur to decrease the degree of spin frustration, which appears to be energetically favorable. Such an effect was described in [68, 69], in which the use of neutron diffraction techniques demonstrated that in frustrated systems the relation between the magnetic and ferroelectric orders could occur, as magnetic interactions were more energetically favorable in case of small changes in the atoms positions. The existence of magnetoelectric coupling was revealed in $Cu_2OCl_2$ containing $OCu_4$ tetrahedral chains similar to those that can be identified in the averievite structure [70].

*3.2.2. Fluctuations of spin configurations.* As was shown above, the O2 ions centering the tetrahedron I (O2Cu3Cu2Cu2Cu2) (figures 4(a), (b) and (c)) and the O3 ions centering the tetrahedron II (O3Cu1Cu2Cu2Cu2) (figures 4(d), (e) and (f)) provide the main contribution to antiferromagnetic components of all the spin interactions in tetrahedra. Therefore, the shift (fluctuation) of the O2 and O3 ions inside the tetrahedral could result in fluctuations of spin configurations. The ability of the of Jahn-Teller coordination of the $Cu^{2+}$ ion to deformation allows shifting of the centering oxygen ions inside the $Cu_4$ tetrahedron, but the shift degree is limited by the tetrahedron size. For the simplicity, we considered shifting of the centering oxygen ions (O2 and O3) only along the *c* axis, which is possibly for the space group *P*3 (figures 8(a)-(c)), although there exist the possibility of the averievite transition to a different symmetry. As was shown in calculations, shifting of the tetrahedron-centering oxygen ion to the tetrahedron vertex (removal from the kagome lattice) increased dramatically the contribution of this ion to the AFM component of interaction along three edges converging in this vertex. In opposite, removal of the oxygen ion from the vertex to the opposite face (approaching the kagome lattice) decreases this contribution until the emergence of the AFM → FM transition. It is important to emphasize that these shifts of the centering oxygen ion in the tetrahedron are not significant for the parameters of magnetic interactions along the edges of the opposite face, included into the kagome lattice. There occurs only the increase of the strength of AFM magnetic couplings in the kagome lattice upon the approaching of centering oxygen ions and, in opposite, the strength decrease upon these ions removal, whereas the AFM → FM transition is not observed (see an example below).

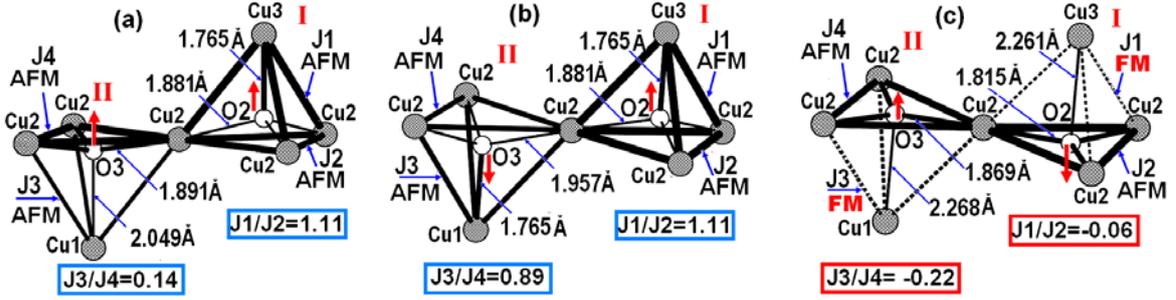

**Figure 8.** Variation of parameters of magnetic interactions $J1$ and $J2$ at shifting of the centering O2 ion in the tetrahedron *I* and $J3$ and $J4$ at shifting of the centering O3 ion in the tetrahedron *II*: (a) shifting of the O2 (by 0.09 Å) and O3 (by 0.13 Å) ions from the centers of the tetrahedra *I* and *II*, respectively, along the 001 direction (local polarization, initial position [24, ICSD-85128]); (b) shifting of the O3 ion by 0.28 Å from the initial position to the tetrahedron vertex Cu1; (c) shifting of the O2 ion by 0.50 Å and the O3 ion by 0.22 Å relatively to the initial position to the kagome plane.

Shifting of the O3 along the 00-1 direction by 0.28 Å (from z = 0.244 to z = 0.210) from the kagome plane to the Cu1 vertex in the tetrahedron *II* increases the degree of competition between AFM magnetic interactions along the edges of the $J3$ and $J4$ tetrahedra from $J3/J4 = 0.14$ to $J3/J4 = 0.89$ (figures 8(a) and (b)). The latter occurs due to the increase (2.4-fold) of the contribution of the O3 ion to the AFM component of the $J3$ interaction and the decrease (just 1.4-fold) of that to the AFM component of the $J4$ interaction between the Cu2 ions in the kagome plane.

In case of centering oxygen ions approaching the centering oxygen ions to the kagome plane, there can be attained the elimination of the $J1$ and $J3$ magnetic couplings of the Cu3 and Cu2 ions, respectively, with the Cu2 ions in the kagome lattice until reorientation of their magnetic moments from AFM to FM (figure 8(c)). According to calculations, at shifting of the O2 ion by 0.50 Å along the 00-1 direction and the O3 ion by 0.22 Å along the 001 directing, the contributions of these ions to the $J1$ and $J3$ magnetic couplings will be transformed from large AFM contributions (j(O2) = -0.071 Å$^{-1}$; j(O3) = -0.028 Å$^{-1}$) to small FM ones (j(O2) = 0.004 Å$^{-1}$; j(O3) = 0.002 Å$^{-1}$). The strengths of AFM $J2$ and $J4$ magnetic couplings between the Cu2 in the kagome plane will increase 1.3- and 1.1-fold, respectively.

Aside from the considered instability of the $J1$ and $J3$ magnetic interactions caused by shifting of the centering ions (O2 and O3, respectively), there exists another factor affecting their parameters. Let us call it "the effect of common edge". The point is, the Cu2 and Cu3 polyhedra are linked through a common O2-O6 edge (figure 4(b)), where as the Cu1 and Cu2 polyhedra are linked through a common O3-O5 edge (figure 4(e)). The oxygen ions forming these edges are intermediate ions determining the parameters of magnetic interactions – $J1$ and $J3$, respectively. Compression and expansion of this edge, just like shifting of ions forming it relatively to the center of coupling of magnetic atoms, determine the direction of their magnetic moments. As we calculated for this copper oxide (Cu$^{2+}$), the expansion of the common edge from 2.46 to 2.75 Å, preserved the coupling AFM character, but decreased the strength of this interaction, whereas the increase of the edge length up to 2.84 Å resulted in reorientation of the magnetic moments: AFM → FM (figures 9(a)-(c)).

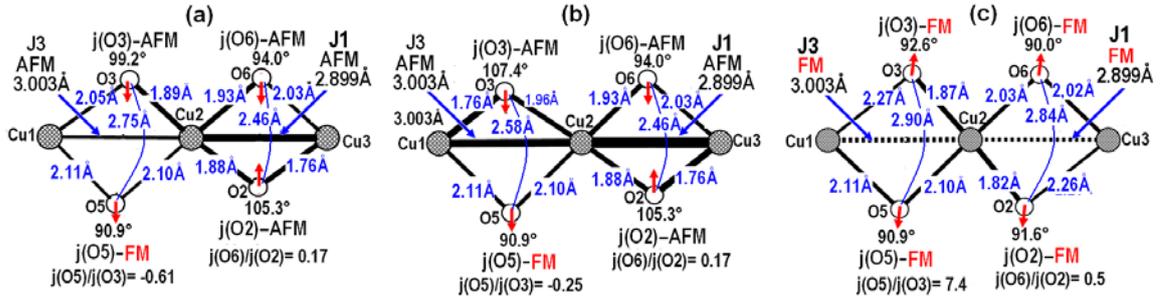

**Figure 9.** Variation of parameters of $J1$ and $J3$ magnetic interactions at compression and expansion of the O3-O5 common edge of Cu1 and Cu2 polyhedra and the common O2-O6 edge of the Cu2 and Cu3 polyhedra (d-f).

To sum up, we have demonstrated the possibility of continuous fluctuation of the spin structure due to vibrations of the centering oxygen ion. One can decrease the frustration degree in the tetrahedron through shifting of the centering oxygen ion, within local limitations imposed by tetrahedra sizes (frameworks). However, the main point consists in the fact that it is impossible to eliminate frustration in this mineral because of the presence of the frustration base – antiferromagnetic kagome lattice.

### 3.3. Ilinskite $NaCu_5O_2(SeO_3)_2Cl_3$

Ilinskite $NaCu_5O_2(SeO_3)_2Cl_3$ [25] crystallizes in the centrosymmetric orthorhombic *Pnma* space group. Magnetic $Cu^{2+}$ ions occupy 4 crystallographically independent sites (Cu1, Cu2, Cu3, and Cu4) and have a characteristic Jahn-Teller distortion of $Cu^{2+}$ coordination polyhedra ($Cu1O_4Cl$ – square pyramid, where d(Cu1-O) = 1.910 - 2.009 Å, d(Cu1-$Cl^{ax}$) = 2.628 Å); $Cu2O_3Cl_2$ – square pyramid, where d(Cu2-O) = 1.964 – 1.975 Å, d(Cu2-$Cl^{eg}$) = 2.211 Å, d(Cu2-$Cl^{ax}$) = 2.779 Å); $Cu3O_3Cl$ – distorted square, где d(Cu3-O) = 1.897 – 1.972 Å, d(Cu3-$Cl^{eg}$) = 2.360 Å); $Cu4O_4Cl$ – square pyramid, where d(Cu4-O) = 1.806 - 2.062 Å, d(Cu4-$Cl^{ax}$) = 2.823 Å). The $CuO_mCl_n$ coordination polyhedra share edges to form tetramers that have 'additional' O1 and O2 atoms as centers (figures 10 (a) and (f)). Also, just like in case of averievite, the base of the crystal structure of illiskite consists of two types of oxocentered $OCu_4$ tetrahedra (tetrahedron I – O1Cu2Cu3Cu4Cu4 and tetrahedron II – O2Cu1Cu3Cu4Cu4) corner-shared into $[O_2Cu_5]^{6+}$ layers (figure 6(c)) in accordance with the kagome lattice motif (figure 7(c)). However, in ilinskite the kagome lattice is slightly corrugated, unlike the regularly flat lattice in averievite. Besides, the layer of corner-sharing $OCu_4$ tetrahedra in averievite is composed of the UDUDUD 6- membered rings of tetrahedra, whereas the layer in ilinskite consists of the UUDUUD and DDUDDU rings. Here the U and D symbols indicate orientation of tetrahedra either up or down relatively to the plane of the layer (figures 6(a) and (c)). Thus, the sublattice of magnetic $Cu^{2+}$ ions in ilinskite, just like in averievite, consists of the corner-sharing $Cu_4$ tetrahedra on the kagome lattice. However, the directions of unbound tetrahedra vertices in them will be different.

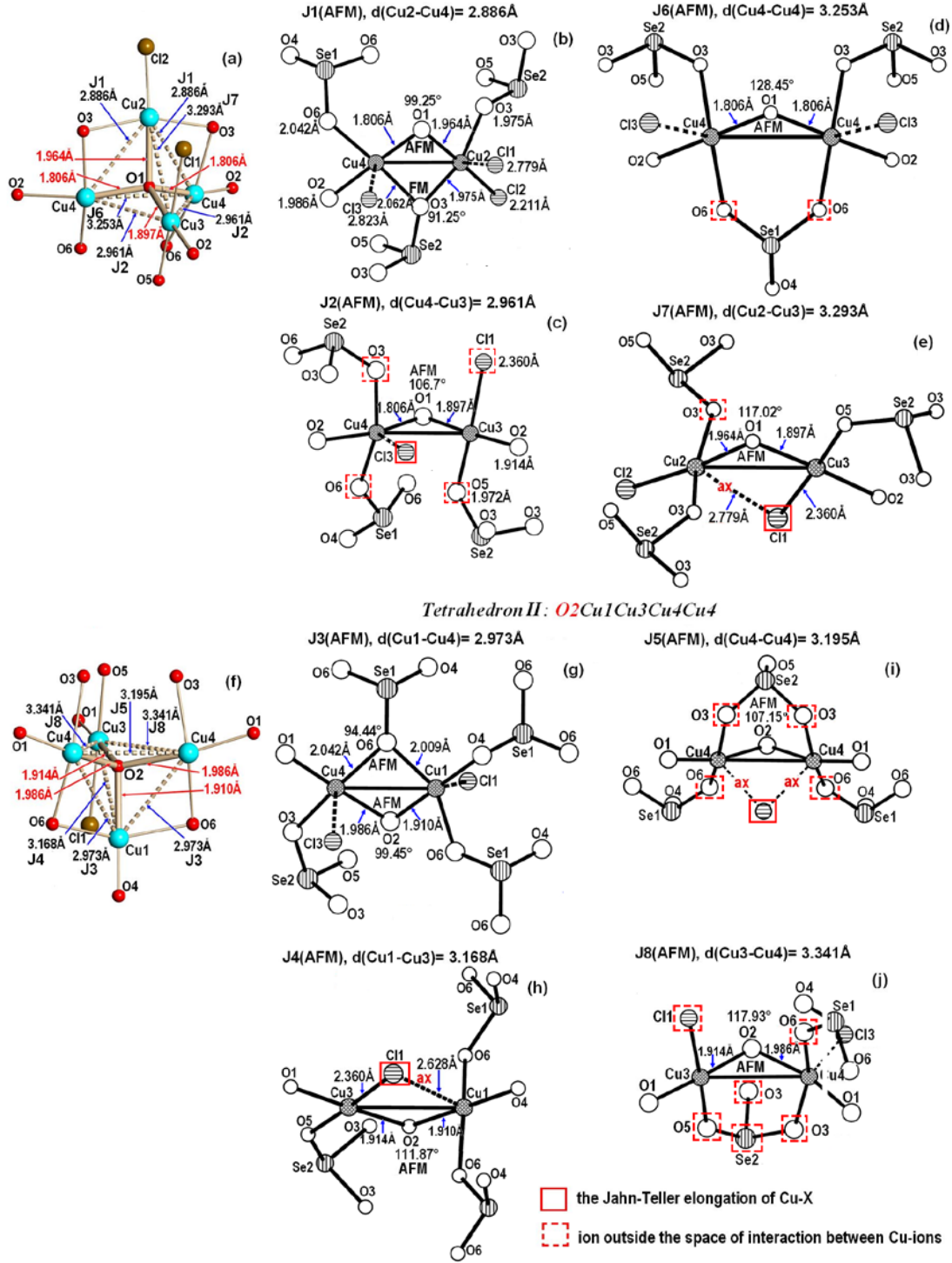

**Figure. 10**. Linking the CuOmCln coordination polyhedra in the tetramers I (a) and II (f) in the fumarolic mineral ilinskite $NaCu_5O_2(SeO_3)_2Cl_3$ [25]. The arrangement of intermediate ions in the local space of AFM $J1$ (b), $J2$ (c), and $J6$ (d) couplings in tetrahedron $I$ and $J3$ (g), $J4$ (h), $J5$ (i), and $J8$ (j) couplings in tetrahedron II.

According to our calculations using the structural data of $NaCu_5O_2(SeO_3)_2Cl_3$ obtained by Krivovichev et al. [25], all the magnetic couplings between the nearest-neighbors in this compounds are antiferromagnetic. The kagome lattice (figure 7(c), table 2) is formed by two types of triangles. In the triangle I (Cu3Cu4Cu4), two AFM $J2$ ($J2$ = -0.0661 Å$^{-1}$, d(Cu3-Cu4) = 2.961 Å) and one AFM $J6$ ($J6$ = -0.1162 Å$^{-1}$, d(Cu4-Cu4) = 3.2531 Å) couplings are strong and compete to each other ($J2/J6$ = 0.57). Here, the AFM $J6$ coupling is predominant in ilinskite. The AFM $J2$ and $J6$ couplings (figures 10(c) and (d)) are formed under effect of the O1 oxygen ion centering the tetrahedron I.

In the triangle II (Cu3Cu4Cu4), one observes a similar picture: two AFM $J8$ ($J8$ = -0.0709 Å$^{-1}$, d(Cu3-Cu4) = 3.341 Å) and one AFM $J5$ ($J5$ = -0.0433 Å$^{-1}$, d(Cu3-Cu4) = 3.195 Å) couplings are strong and compete to each other ($J5/J8$ = 0.61). The AFM $J5$ and $J8$ couplings (figures 10(i) and (j)) are formed under effect of the O2 oxygen ion centering the tetrahedron II.

In addition to the competition in triangles, there exists the competition of the nearest-neighbor $J5$ and $J6$ couplings with the next-nearest-neighbor $J^{5+6}$ ($J^{5+6}/J5$ = 0.79; $J^{5+6}/J6$ = 0.29) ones in linear chains along the triangles sides parallel to the $b$ axis in the kagome plane. The next-nearest-neighbor AFM $J_{Cu4}^{2+8}$ and FM $J_{Cu3}^{2+8}$ couplings along two other triangles sides are weak and cannot compete to the strong AFM $J2$ ($J_{Cu4}^{2+8}/J2$ = 0.04; $J_{Cu4}^{2+8}/J8$ = 0.03,) and $J8$ ($J_{Cu3}^{2+8}/J2$ = -0.02; $J_{Cu3}^{2+8}/J8$ = -0.01) couplings. The second-neighbor couplings of the $J_2$ type ($J9$, $J10$ and $J11$) and the third-neighbor ones of the $J_3$ type ($J12$ and $J_b^{3-3}$) (figures 3(c), 6(c), 7(c); table 2) are absent in the kagome plane ($J9$ = 0, $J10$ = 0, $J11$ = 0, $J12$ = 0, $J_b^{3-3}$ = 0). Along with the presence of a strong frustrations of the AFM nearest-neighbor couplings in the triangles I ($J2$ and $J6$) and II ($J5$ and $J8$), this could indicate to the existence of a spin liquid (-0.03 ≤$J_2/J_1$≤ 0.045), according to [61]) in case of an isolated kagome lattice.

However, in both ilinskite and averievite, the Cu2 and Cu1 ions located above and below the lattice supplement the AFM spin-frustrated triangles of the kagome lattice until AFM spin-frustrated tetrahedra. As in the tetrahedron I, couplings of the Cu2 ion with the Cu4 (AFM $J1$: $J1/J6$ = 0.37, d(Cu2-Cu4) = 2.886 Å) and Cu3 (AFM $J7$: $J7/J6$ = 0.62, d(Cu2-Cu3) = 3.293 Å) ions in the tetrahedron II, and couplings of the Cu1 ion with the Cu4 (AFM $J3$: $J3/J8$ = 0.53, d(Cu1-Cu4) = 2.973 Å) and Cu3 ($J4$: $J4/J8$ = -0.93, d(Cu1-Cu3) = 3.168 Å) ions are rather strong. Aside from the considered strong AFM couplings, the layer contains two more strong AFM $J15$ ($J15$ = -0.0659 Å$^{-1}$, d(Cu2-Cu4) = 5.379 Å) and $J16$ ($J16$ = -0.0703 Å$^{-1}$, d(Cu1-Cu4) = 5.460 Å) couplings between the nearest tetrahedra (figure 6(c), table 2). They compete to the couplings in the kagome lattice in AFM $J15J1J5$ triangles ($J1/J15$ = 0.66, $J5/J15$ = 0.66) and $J16J3J6$ ($J3/J16$ = 0.54, $J6/J16$ = 1.65). It remains unclear what effect can be provided by these additional frustrated AFM couplings on the possibility of the emergence of the spin liquid state.

All the couplings ($J19$–$J24$) between the layers at distances in the range 4.78 Å – 7.41 Å, except one, are very weak (table 2). They are 20–58-fold weaker than the predominant $J6$ coupling. Only the AFM $J23$ ($J23$ = -0.0108 Å$^{-1}$, d(Cu2-Cu4) = 7.242 Å) couplings are stronger than others ($J23/J6$= 11) and could weakly compete to the $J5$ couplings from the kagome plane in the AFM $J23J5J23$ triangles ($J23/J5$ = 0.25).

Kovrugin et al. [71] studied again the crystal structure of $NaCu_5O_2(SeO_3)_2Cl_3$ and determined that of its K-analog. Note that all the figures (figures 6(c) and (d), 7(c) and (d), and 9) and markings of atoms and magnetic $J$n couplings (except interatom distances and angles values) are similar for the samples of the fumarolic mineral and synthetic ilinskite $NaCu_5O_2(SeO_3)_2Cl_3$ and its K-analog $KCu_5O_2(SeO_3)_2Cl_3$.

Differences in the structural data of $NaCu_5O_2(SeO_3)_2Cl_3$ obtained in [25] and [71] by means of X-ray single-crystal diffraction (the refinement converged to the residual factor (R) values R = 0.044 and 0.049, respectively) are insignificant. Nevertheless, we have calculated respective parameters of magnetic couplings in the synthetic $NaCu_5O_2(SeO_3)_2Cl_3$ on the basis of new structural data [71] (marked by asterisk). It turned out that all the magnetic couplings in both samples (Na-synthetic and fumarolic mineral ilinskite) were identical in sign, but four respective $J1(J1*)$ and $J6(J6*)$ couplings in the tetrahedron *I* and $J3(J3*)$ and $J5(J5*)$ ones in the tetrahedron *II* differ noticeably in strength ($J1*/J1 = 0.66$, $J6*/J6 = 0.77$, $J3*/J3 = 1.33$ and $J5*/J5 = 1.86$). The latter is related to the fact that at interaction of magnetic ions at short distances even small shifts of the intermediate O1 and O2 ions in the local space of these interactions induce substantial strength changes. The parameters of magnetic couplings between $Cu^{2+}$ ions at long distances differ insignificantly for these two samples.

In addition, we have calculated the parameters of magnetic $Jn$ couplings in the K-analog based on the structural data [71]. It turned out that all the $Jn$ parameters in synthetic $KCu_5O_2(SeO_3)_2Cl_3$ are similar in sign and strength to respective parameters in synthetic $NaCu_5O_2(SeO_3)_2Cl_3$.

To sum up, according to our calculations, the magnetic structures of two samples of synthetic and fumarolic mineral ilinskite $NaCu_5O_2(SeO_3)_2Cl_3$ and its K-analog, just like of $Cu_5O_2(VO_4)_2(Cu^+Cl)$, consist of AFM spin-frustrated layers of the corner-sharing $Cu_4$ tetrahedra *I* and *II* on the kagome lattice (figures 6(c) and (d), 7(c)). However, the directions of unbound tetrahedra vertices are different in them. The emergence of the AFM character of couplings in the tetrahedra *I* and *II* is caused by the O1 and O2 oxygen ions, respectively, centering these tetrahedra. The parameters of magnetic couplings in the kagome lattice for all the samples of ilinskite, just like those of averievite, meet the criteria of the existence of the spin liquid in the case of an isolated kagome lattice [58-65].

It is worth mentioning that the magnetic model for $KCu_5O_2(SeO_3)_2Cl_3$ built in [72] is different from ours. The magnetic exchange couplings for this model were obtained from first-principles calculations within the framework of the density functional theory (DFT) with the generalized gradient approximation (GGA) for the exchange-correlation potential. The model comprises weakly coupled spin ladders (tubes) having a complex topology formed upon fragmentation of the tetrahedral network. This fragmentation is rooted in the nontrivial effect of the $SeO_3$ groups that render the Cu–O–Cu super-exchange strongly ferromagnetic even at bridging angles exceeding 110°. The differences are mainly related to three interactions ($J2$ (d(Cu3-Cu4) = 3.148Å), $J5$ (d(Cu4-Cu4) = 3.168 Å) and $J8$ (d(Cu3-Cu4) = 3.173 Å)), which are, according to our calculations, strong AFM couplings, whereas in [72] they are marked as $J3$, $J4$, and $J5$ and defined as strong FM couplings. That is why figure 10 showing the arrangement of intermediate ions in the local interaction space additionally demonstrates linking of $Cu^{2+}$ ions with $SeO_3$-groups. In all the $J2$, $J5$, and $J8$ interactions, the oxygen ions from $SeO_3$-groups are outside the local interaction space and, therefore (according to our concept), do not contribute to the formation of magnetic couplings.

However, the above disagreement can be examined from another aspect. Because of the effect of a lone electron pair, the $Se^{4+}$ ions are characterized with an "umbrella" coordination in $SeO_3$-groups. In view of this, $SeO_3$-groups are very mobile and, could, under effect of temperature or pressure, easily rotate or shift. As a result, the O3, O5, and O6 oxygen atoms from $SeO_3$-groups located near the boundaries of local spaces of the *J*2, *J*5, and *J*8 interactions (figure 10) could enter these spaces and induce the spin reorientation of the AFM–FM type.

Let us consider the strong AFM *J*8 (*J*8 = -0.0706 Å$^{-1}$, d(Cu3-Cu4) = 3.173 Å) coupling (figure 10(j)). It is marked as *J*5 and defined as a strong FM coupling in [72]. The contribution to formation of the AFM character of this coupling is provided by one O2 oxygen ion centering the tetrahedron II (O2Cu1Cu3Cu4Cu4). There are no other intermediate ions in the local space of this coupling. According to our concept, this coupling could become a strong FM one, if O5 ions (from the Se2O$_3$-group) and/or O6 ions (from the Se1O$_3$-group) enter the local space of this coupling. The latter will take place, if SeO$_3$-groups approach the Cu3-Cu4 bond line. As can be roughly described, within the frames of the initial space group *Pnma*, shifting of the O5 ion (from the initial value z(O5) = -0.2102 to z(O5) = -0.1240) along the *z* axis by 0.91 Å will result in the emergence of a strong contribution *j*(O5) = 0.1365 to the FM component of the *J*8 couplings exceeding the AFM contribution of *j*(O2). Finally, the *J*8 coupling will transform to the ferromagnetic one (*J*8 = 0.0659 Å$^{-1}$).

Earlier [73, 74] in studies of the stereochemical role of the pair, we demonstrated that presence of ions having the lone pair of electrons in the crystal structure already has a potential for phase transitions, including magnetic ones. The lone pair of electrons is responsible for loose parts in the structure and can easily change its position, thus providing the possibility for atoms to shift under effect of external forces. Structural phase transitions accompanied with magnetic transitions can occur as within the frames of the same space group as with symmetry changes, for example, replacement of a noncentrosymmetrical space group by a symmetrical one and vice versa. In our opinion, the Na- and K-ilinskite minerals are characterized with a versatility of magnetic transitions not less rich than BiFeO$_3$: in the latter case, as we showed in [74, 75], the lone pair of electrons of the trivalent bismuth has an important role.

### 3.4. Avdoninite $K_2Cu_5Cl_8(OH)_4 \cdot 2H_2O$

The crystal structure of the mineral avdoninite [27] is similar to that of its synthetic analog $K_2Cu_5Cl_8(OH)_4 \cdot 2H_2O$ reported by Kahlenberg [26]. To calculate the parameters of magnetic couplings, we used the data on the synthetic analog of avdoninite ([26], ICSD-55096). Avdoninite is monoclinic, has a centrosymmetric space group *P*2$_1$/*c*, *a* = 11.6424(1), *b* = 6.5639(4), *c* = 11.7710(10) Å, β = 91.09(1)°, *Z* = 2. The crystal structure of this mineral is based on sheets of copper-oxo-chloride complexes ([Cu$_5$Cl$_8$(OH)$_4$]$^{2-}$) perpendicular to the (100) direction. The K+ cation and H$_2$O molecules are interlayers.

Magnetic Cu$^{2+}$ ions occupy 3 crystallographically independent sites (Cu1, Cu2, and Cu3) and have a characteristic Jahn–Teller distortion of Cu$^{2+}$ coordination polyhedra (figure 10). The Cu1 and Cu3 coordination polyhedra can be described as strongly distorted octahedra. The Cu1 atom has [2O + 2Cl] square *trans*-coordination, where d(Cu1-O2) = 1.963 Å and d(Cu1-Cl1) = 2.329 Å, which is added by two long Cu1–Cl2$^{ax}$ (d(Cu1-Cl2$^{ax}$ = 2.876 Å)) bonds to form distorted octahedral coordination. The Cu3 atom has [3O + 1Cl] square coordination, where d(Cu3-O) = 1.973 – 2.010 Å and d(Cu3-Cl2) = 2.298 Å and, in addition, two long Cu3–Cl1$^{ax}$ (d(Cu3-Cl1$^{ax}$ = 2.662 and 2.711 Å)) bonds. The Cu2 atom is a distorted Cu2O$_2$Cl$_3$ square pyramid, where two inner Cu2-O and two inner Cu2-Cl bond distances within the slightly distorted square about Cu2 are in the range between 1.992 and 2.22 Å and additionally one Cl2 ion at a long distance (d(Cu2-Cl2) =2.884 Å) in the pyramid vertex.

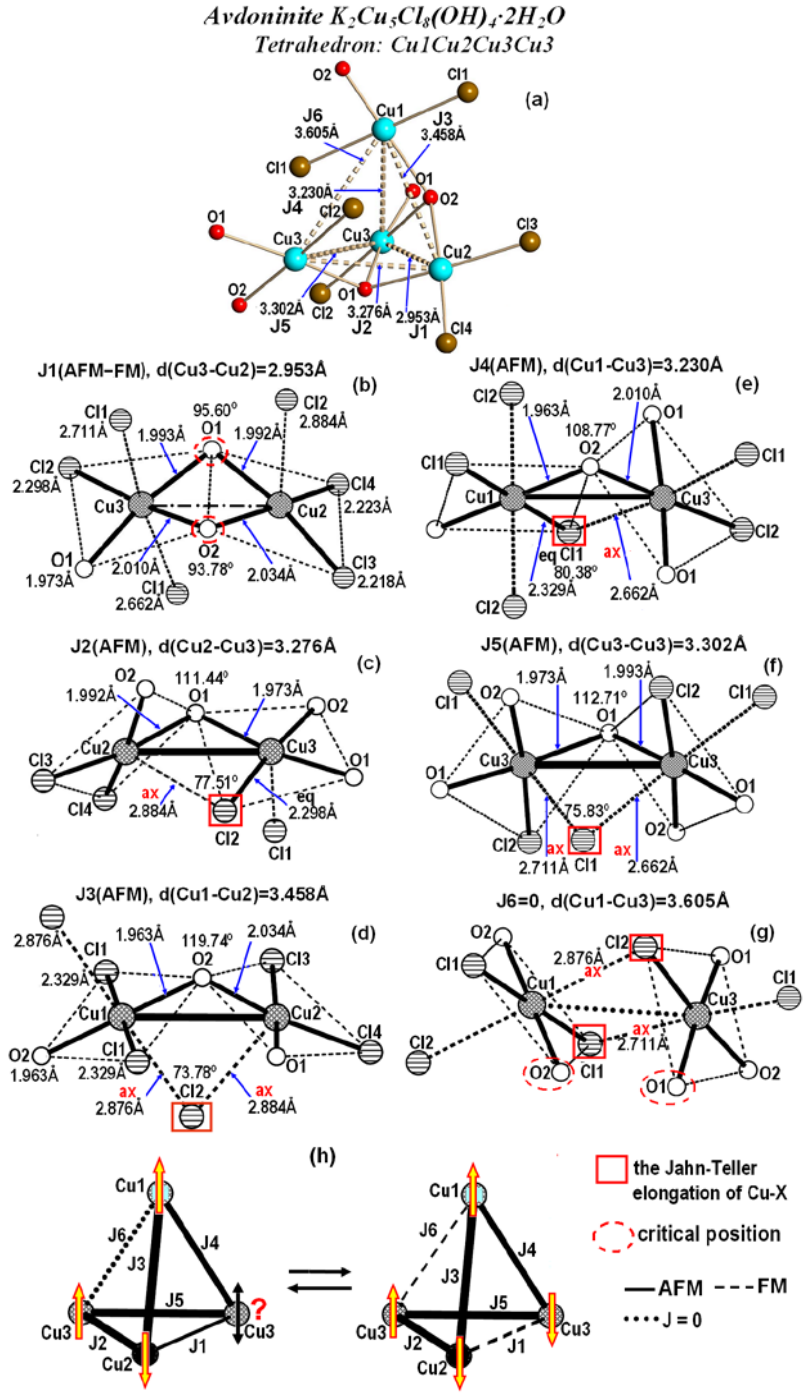

**Figure 11.** Linking of the CuOmCln coordination polyhedra into tetramer in avdoninite $K_2Cu_5Cl_8(OH)_4 \cdot 2H_2O$ (a); the arrangement of intermediate ions in the local space of $J1 - J6$ interactions in tetrahedron $Cu_4$ (b-g) and possible spin reorientation and short-range AFM order in tetrahedron (h).

Let us consider magnetic characteristics of couplings in avdoninite in comparison with respective couplings in averievite and illiskite. The crystal sublattice of magnetic $Cu^{2+}$ ions in avdoninite $K_2Cu_5Cl_8(OH)_4·2H_2O$, just like in averievite $Cu_5O_2(VO_4)_2(Cu^+Cl)$, consists of layers of corner-sharing tetrahedra located on the distorted kagome lattice, so that unshared tetrahedra vertices (Cu2) are directed to opposite sides relatively to this lattice. These layers are perpendicular to the *a* axis and repeat at distances equal to the *a* parameter. Unlike averievite and illiskite, in avdoninite the crystal sublattice of magnetic $Cu^{2+}$ ions contains just one type of tetrahedra formed by copper ions (Cu1Cu2Cu3Cu3). However, the main reason of the difference in magnetic characteristics of these minerals consists in the absence of the oxygen ions in the centers of $Cu_4$ tetrahedra in averievite (table 2, figure 11(a)). Below we will examine it in more detail.

The oxygen ions centering tetrahedra in averievite and illiskite make not equal, but substantial AFM contributions to each of six couplings along the tetrahedra edges, which serves as a reason of their AFM character and, as a result, their frustration. All the couplings along six tetrahedra edges in averievite are very different. The AFM *J*3 (figure 6(e) and (f), 11(d)) and AFM *J*2 (figure 6(e) and (f), 11(c)) couplings of the Cu2 ions located above and below the kagome lattice with the Cu1 and Cu3 ions from this lattice are strong. The AFM *J*3 ($J3$ = -0.0665 Å$^{-1}$, d(Cu2-Cu1) = 3.458 Å) coupling, which is predominant in the structure, is formed under effect of the O2 ion. The AFM *J*2 ($J2/J3$ = 0.79, d(Cu2-Cu3) = 3.276 Å) coupling is formed under effect of the O1 ion. The third AFM *J*1 ($J1/J3$ = 0.27, d(Cu2-Cu3) = 2.953 Å) coupling of the Cu2 ion with the Cu3 ion from the kagome plane is very weak. Small AFM contributions to its formation are provided by two oxygen ions: O1 and O2 (figures 6(e) and (f), 11(b)). Moreover, at insignificant shifts of the O1 and O2 ions from the Cu2-Cu3 bond line and their leaving of the local space, the *J*1 transition from the AFM to the FM state of the AFM $J1 \rightarrow 0 \rightarrow$ FM $J1$ type is possible.

Inequality of characteristics of magnetic couplings in the kagome lattice (figures. 7(e) and (f)) is also caused by the absence of the oxygen ion in the tetrahedron center. In small Cu1Cu3Cu3 triangles of the kagome lattice, two AFM $J5_1$ ($J5_1/J3$ = 0.83, d(Cu3-Cu3) = 3.302 Å) (figure 11(f)) and AFM $J4_1$ ($J4_1/J3$ = 0.70, d(Cu1-Cu3) = 3.230 Å) (figure. 11(e)) couplings are strong, whereas the third $J6_1$ (d(Cu1-Cu3) = 3.605 Å) coupling (figure. 11(g)) is equal to 0. The local space of this coupling contains just two ions (Cl1 and Cl2), whose bonding with copper ions has a Jahn-Teller elongation, so that they cannot contribute to the magnetic coupling. At the same time, the O1 and O2 oxygen ions are present near the local space of the $J6_1$ coupling: they are able, under effect of temperature or pressure, shift in parallel to the Cu1-Cu3 bond line to the center, enter the local interaction space, and initiate the emergence of comparatively small contributions to the FM component of the $J6_1$ interaction. A short-range AFM order in the tetrahedron (figure 11(h) could emerge as a result of simultaneous reorientation of spins of two *J*1 and $J6_1$ couplings into the FM state along the Cu2-Cu3 and Cu1-Cu3 edges.

Unlike averievite and illiskite, in honeycombs lattice of avdoninite, the kagome couplings at long distances along sides of large triangles (*J*7 (d(Cu1-Cu3) = 5.196 Å), *J*8' (d(Cu3-Cu3) = 5.997 Å) and *J*9 (d(Cu1-Cu3) = 6.287 Å)) are not equal to 0. They are weak AFM couplings (9–13-fold weaker than the predominant *J*3 coupling) and compete to each other (figure 7(e), table 2). The strongest interlayer coupling (*J*20 (d(Cu2-Cu2) = 7.999 Å)) is 7-fold weaker than the *J*3 coupling predominant in the layer (figure 7(f)).

Finally, let us demonstrate on the example of avdoninite, what will be the result of transformation of the magnetic structure of the pyrochlore layer existing in averievite, of one removes from tetrahedra oxygen ions centering them. Inequality of the strength of the nearest-neighbor couplings in tetrahedra upon removal of oxygen ions from their centers, which is clearly expressed in avdoninite as elimination of the $J6_1$ coupling, changes the general picture dramatically.

First, the AFM tetrahedron transforms into an open tetrahedron or two frustrated AFM $J1J2J5_1$ (on the Cu3Cu2Cu3 triangle) and $J1J3_1J4_1$ (one the Cu1Cu2Cu3 triangle) triangles with a shared $J1$ edge (Cu2-Cu3) (figure 6(e)). Such a configuration is usually called "butterfly". Second, there occurs an elimination of the kagome plane (figure 7(e). The structure of a layer in the *bc* plane transforms into an openwork curled net with large cells weaved from corner-sharing open AFM spin-frustrated tetrahedra (butterflies) (figure 6(e)).

In the formed net, triangles of the Cu3Cu2Cu3 type are linked by common Cu3 vertices into chains spread along the *b* axis (figure 6(e)). The nearest-neighbor AFM $J1$ ($J1/J5_1 = 0.33$), $J2$ ($J2/J5_1 = 0.095$), and $J5_1$ ($J5_1 = -0.0553$ Å$^{-1}$) couplings along the edges of these triangles compete to each other. There is no another competition in this chain due to negligible values of the next-nearest-neighbor couplings, also in the linear chain along the *b* axis with the nearest-neighbor AFM $J5_1$ and the next-nearest-neighbor AFM $J5_2$ ($J5_2/J5_1 = 0.03$) intrachain couplings. These chains of AFM spin-frustrated triangles are linked to each other into a network through triangles of a different type (Cu1Cu2Cu3) from an open tetrahedron, which are linked pairwise through a common Cu1 vertex (figure 6(e)).. There also exists a competition between the nearest-neighbor AFM $J1$ ($J1/J3_1 = 0.27$), AFM $J4_1$ ($J4_1/J3_1 = 0.70$), and AFM $J3_1$ ($J3_1 = -0.0665$ Å$^{-1}$) couplings along the edges of these triangles. Besides, in this pair of tetrahedra, there exist additional competitions between the nearest-neighbor $J3_1$ and the next-nearest-neighbor AFM $J3_2$ ($J3_2/J3_1 = 0.45$) couplings, as well as between the nearest-neighbor $J4_1$ and the next-nearest-neighbor AFM $J4_2$ ($J4_2/J4_1 = 0.67$) couplings (figure 6(f)). Thus, we demonstrated that in avdoninite the compliance of the magnetic structure with the crystal structure of the sublattice of $Cu^{2+}$ magnetic ions is disrupted.

### *3.5. Structural-magnetic models in search of new frustrated magnetic materials.*

Let us define the structure of the compound, in building of which the sign and strength of magnetic interactions are calculated by the crystal chemistry method we developed [32-34], as the structural-magnetic model. Such a structural-magnetic model is based on crystal chemistry parameters (crystal structure and ions sign and strength). This model is characterized with (1) sign and strength of magnetic couplings; (2) dimensions of the magnetic structure, which not always coincide with those of the crystal structure; (3) presence of magnetic frustrations on specific geometric configurations; and (4) possibility of reorientation of magnetic moments (transition of the AFM – FM type) at shifts of intermediate ions localized in critical positions. The structural-magnetic models enable one to reveal main correlation relationships between the compounds structures and magnetic properties and to determine, on their basis, the crystal chemistry criteria for targeted search of new functional magnetics in the Inorganic Crystal Structure Data (ICSD) database. In our search of frustrated magnetics, a special interest is related to structures including a low-dimensional sublattice in the form of chains and layers composed of corner-sharing oxocentered $OCu_4$ tetrahedra on the kagome lattice and kagome lattice of $Cu^{2+}$ ions.

In [23], we determined the structural-magnetic models of $Cu_3Mo_2O_9$ and the volcanic mineral kamchatkite ($KCu_3OCl(SO_4)_2$) containing spin-frustrated pyrochlore chains and compared them with magnetic structures determined in the experiment. It turned out that the results of calculations of the parameters of magnetic couplings for $Cu_3Mo_2O_9$ by the crystal chemistry method were in good agreement with the experimental data [52]. The found discrepancies are not of a principal character while observed mainly in the values of the strength of magnetic couplings at short distances. This is related to the fact that at interaction of magnetic ions located at short distances even slight shifts of intermediate ions in the local interaction space produce significant changes in the strength of magnetic couplings. Besides, we demonstrated that, unlike the crystal chemistry method, experimental methods were not so sensitive to local changes in the magnetic coupling strength. They do not distinguish between couplings that are geometrically close, but crystallographically nonequivalent, for example, $J1_1$ and $J2_1$ for $Cu_3Mo_2O_9$ (Fig.4, Table 1 in [23]). The reason of incompliance of calculated and experimental data could also consist in determination of the crystal structure and magnetic couplings parameters at different temperatures and by different methods.

**Table 3.** Crystallographic characteristics and parameters of magnetic couplings ($J_n$) calculated by crystal chemistry method and *ab initio* $GGa+U$ in tetrahedral minerals: averievite $Cu_5O_2(VO_4)_2(Cu^+Cl)$ and $Cu_5O_2(VO_4)_2(CsCl)$ at 400 K

| Crystallographic and magnetic parameters | $Cu_5O_2(VO_4)_2(Cu^+Cl)$ [24] Space group *P*3 (N143) $a = b = 6.375$, $c = 8.399$ Å $\alpha = \beta = 90°$, $\gamma = 120°$, Z=1 Method[a] – XDS; R-value[b] = 0.052 | | $Cu_5O_2(VO_4)_2(CsCl)$ 400K [78] Space group *P*-3*m* (N164) $a = b = 6.3693$, $c = 8.3758$ Å $\alpha = \beta = 90°$, $\gamma = 120°$, Z = 1 Method[a]–S-XPD; R-exp[b] = 5.96 | $Cu_5O_2(VO_4)_2(CsCl)$ 400K [78] Shifting of O2 from z(O2) = 0.95 to z(O2) = 0.97 |
|---|---|---|---|---|
| | Tetrahedron I O2Cu3Cu2Cu2Cu2 | Tetrahedron II O3Cu1Cu2Cu2Cu2 | Tetrahedron I O2Cu1Cu2Cu2Cu2 | Tetrahedron I O2Cu1Cu2Cu2Cu2 |
| $Cu_k$-$Cu_k$ d(Cu-Cu) (Å) | Cu2-Cu2 3.144 | Cu2-Cu2 3.233 | $Cu_k$-$Cu_k$ 3.185 | $Cu_k$-$Cu_k$ 3.185 |
| $J_k$[c] (Å$^{-1}$) | $J2 = -0.0741$ | $J4 = -0.0801$ | $J1 = -0.0771$ (AFM) | $J1 = -0.0881$ |
| $J_k$ (K) ($GGa+U$) [80] | - | - | $J1 = 203$ K (AFM) | - |
| $j(X)$[d] (Å$^{-1}$) $\Delta h(X)$[e] Å, $l_n'/l_n$[f], $CuXCu$[g] | $j(O2)$: -0.0741 -0.295, 1.1, 105.3° | $j(O3)$: -0.0801 -0.419, 1.0, 117.5° | $j(O2)$: -0.0771 -0.391, 1.0, 115.3° | $j(O2)$: -0.0881 -0.447, 1.0, 118.2° |
| $Cu_k$-$Cu_h$ d(Cu-Cu) (Å) | Cu2-Cu3 2.899 | Cu1-Cu2 3.003 | $Cu_k$-$Cu_h$ 2.919 | $Cu_k$-$Cu_h$ 2.919 |
| $J_h$[c] (Å$^{-1}$) | $J1 = -0.0822$ | $J3 = -0.0111$ | $J2 = -0.0392$ (AFM) | $J2 = -0.0151$ |
| $J_h$ (K) ($GGa+U$) [78] | - | - | $J2 = 35$ K (AFM) | - |
| $j(X)$[d] (Å$^{-1}$) $\Delta h(X)$[e] Å, $l_n'/l_n$[f], $CuXCu$[g] | $j(O2)$: -0.0705 (-0.295, 1.1, 105.3°) | $j(O3)$: -0.0282 (-0.126, 1.2, 99.2°) | $j(O2)$: -0.0551 (-0.234, 1.0, 102.8°) | $j(O2)$: -0.0309 (-0.130, 1.2, 97.8°) |
| $j(X)$[d] (Å$^{-1}$) $\Delta h(X)$[e] Å, $l_n'/l_n$[f], $CuXCu$[g] | $j(O6)$: -0.0117 (-0.049, 1.1, 94.0°) | $j(O5)$: 0.0171 (0.077, 1.0, 90.9°) | $j(O3)$: 0.0159 (0.067, 1.07, 89.65°) | $j(O3)$: 0.0159 (0.067, 1.1, 89.6°) |
| $J_k/J_h$ (crystal chemistry method) | $J2/J1 = 0.90$ | $J4/J3 = 7$ | $J1/J2 = 2$ | $J1/J2 = 6$ |
| $J_k/J_h$ [78]; ($GGa+U$ calculations) | - | - | $J1/J2 = 6$ | |

[a] Method: XDS – X-ray diffraction from single crystal; S-XPD - Synchrotron X-ray powder diffraction.
[b] The refinement converged to the residual factor (*R*) values.
[c] $J_n$ in Å$^{-1}$ ($J_n$ (meV) = $J_n$ (Å$^{-1}$)×K, where scaling factors $K_{middle}$ = 74) – the magnetic couplings ($J_n<0$ - AFM, $J_n>0$ – FM).
[d] $j(X)$ – contributions of the intermediate X ion into the AFM ($j(X) <0$) and FM ($j(X)>0$) components of the $J_n$ coupling
[e] $\Delta h(X)$ – the degree of overlapping of the local space between magnetic ions by the intermediate ion X.
[f] $l_n'/l_n$ – the asymmetry of position of the intermediate X ion relatively to the middle of the $Cu_i$–$Cu_j$ bond line.
[g] $Cu_iXCu_j$ – bonding angle.

Unlike the case of $Cu_3Mo_2O_9$, rather insufficient data are available in the literature on the magnetic structure of kamchatkite $KCu_3OCl(SO_4)_2$. However, the conclusion based on experimental measurements made in [76, 77] on frustration and one-dimensional character of this compound is in full agreement with the results of our studies.

Recently, already after submission of this work, there emerged a very good opportunity to compare the results of our calculations with those by Botana al. [78] for properties of similar averievite obtained through *ab initio* calculations along with susceptibility and specific heat measurements. First in [78], the exchange coupling constants of the two nearest neighbor (NN) couplings were provided for the centrosymmetric trigonal at 400 K sample of $Cu_5O_2(VO_4)_2(CsCl)$:

$J$1 connecting Cu-kagome ions and $J$2 between Cu-kagome and Cu-honeycomb ions. Thereafter, the structural data were provided in [78] in addition.

We calculated the parameters of magnetic couplings for the trigonal modification at 400 K (table 3) using the structural data of [78] and came to similar conclusion that all the couplings in $Cu_4$ tetrahedra are antiferromagnetic. The values of couplings between copper ions ($Cu_k$) located at shortest distances in the kagome plane ($J_k$) are stronger than those ($J_h$) with copper ions ($Cu_h$) located above and below it. However, the value of the magnetic couplings strengths ratio ($J_k/J_h$) in our case appeared to be significantly smaller ($J_k/J_h = 2$) than in [78], in which it is equal to 6. One can achieve the same result ($J_k/J_h = 6$), if in this sample ($Cu_5O_2(VO_4)_2(CsCl)$ at 400 K) one shifts O2 ions to the 001 direction (from the initial value z(O2) = 0.9502 to z(O2) = 0.9700) by just 0.0198 (0.166 Å). Here, according to our calculations, the $J_h$ strength will decrease 2.6-fold while the $J_k$ strength will increase insignificantly, and this shift will not virtually affect the lengths of Cu-O bonds. They will remain in the same range (1.850 – 2.105) as in [78].

Let us consider another example. The noncentrosymmetric trigonal mineral averievite $Cu_5O_2(VO_4)_2(Cu^+Cl)$ we investigated in the present work contains two types of oxocentered $OCu_4$ tetrahedra characterized with rather different $J_k/J_h$ ratios (Table 3). In the tetrahedron II $J_k/J_h = 7$, whereas in the tetrahedron I $J_k/J_h = 0.9$. Analysis of the Table 3 data shows that the value of magnetic couplings in the kagome plane ($J_k$) is stable and changes insignificantly from -0.0741 Å$^{-1}$ to -0.0881 Å$^{-1}$, while the $J_h$ value varies in a broad (from -0.0111 Å$^{-1}$ to -0.0822 Å$^{-1}$) range depending mainly on the position of the tetrahedron-centering oxygen ion.

**Table 4** Exchange coupling constants ($J_n$) for ZnCu3(OH)6Cl2 (herbertsmithite) we calculated by the crystal chemistry method (unit: Å$^{-1}$, AFM<0) and determined in [79] from total energies of nine different spin configurations. Energies were calculated with GGA+U functional at $U$ = 6 eV, $J$ = 1 eV (unit: K, AFM>0)

| Herbertsmithite $ZnCu_3(OH)_6Cl_2$ [8] (Data for ICSD – 425834) | | |
|---|---|---|
| Space group R -3 mH (N166); $a = b = 6.834$, $c = 14.075$ Å; $\alpha = \beta = 90°$, $\gamma = 120°$, Z =3 | | |
| Method - X-ray diffraction from single crystal (296 K); R-value = 0.0118 | | |
| Kagome plane couplings | | |
| d(Cu-Cu) (Å) | | 3.417 |
| $J1$ (Å$^{-1}$) | -0.0670 (AFM) | 182.4 K (AFM) [79] |
| d(Cu-Cu) (Å) | | 5.918 |
| $J2$ (Å$^{-1}$) | -0.0108 (AFM) | 3.4 K (AFM) [79] |
| ($J2/J1$) | 0.16 | 0.02 |
| d(Cu-Cu) (Å) | | 6.834 |
| $J3(J_d)$ (Å$^{-1}$) | 0.0018 (FM) | -0.4 K (FM) [79] |
| ($J3(J_d)/J1$) | -0.03 | -0.002 |
| d(Cu-Cu) (Å) | | 6.834 |
| $J1_2$ (Å$^{-1}$) | -0.0300 (AFM) ↔ 0.0178 (FM) | - |
| ($J1_2/J1$) | 0.45 ↔ - 0.26 | - |
| Interplane couplings | | |
| d(Cu-Cu) (Å) | | 5.090 |
| $J4$ (Å$^{-1}$) | -0.0020 (AFM) | 5.3 K (AFM) [79] |
| ($J4/J1$) | 0.03 | 0.03 |
| d(Cu-Cu) (Å) | | 6.130 |
| $J5$ (Å$^{-1}$) | 0.0032 (FM) | -1.5 K (FM) [79] |
| ($J5/J1$) | -0.05 | -0.01 |
| d(Cu-Cu) (Å) | | 6.130 |
| $J5'$ (Å$^{-1}$) | $J5'$ = -0.0012 (AFM) | - |
| ($J5'/J1$) | 0.02 | - |

Further we plan to search for potential spin liquids in the Inorganic Crystal Structure Data (ICSD) database on the basis of the structural-magnetic model of herbertsmithite [ZnCu3(OH)6Cl2]. In view of this, we used the crystal chemistry method to calculate parameters of magnetic couplings in herbertsmithite using the structural data of the sample provided in [8] and compared our results with those of *ab initio* calculations performed in [79] (table 4). There are no principal differences between the data obtained by different methods. In both cases, the AFM nearest neighbor couplings within the kagome planes are predominant. The parameters of $J1$, $J2$, and $J3$ couplings (figure 3(c)) in herbertsmithite are in the range $0.3<J_2 \approx J_3<0.7$ [62-65], which determines the possibility of existence of the chiral QSL state. Couplings between the kagome planes are weak.

To sum up, we have demonstrated that the developed crystal chemistry method enables one to obtain adequate results and can be applied in building structural-magnetic models for search of promising magnetics on the basis of the compounds crystal structure data.

## 4. Conclusions

We have determined the parameters (sign and strength) of magnetic couplings in three volcanic minerals: averievite $Cu_5O_2(VO_4)_2(CuCl)$, ilinskite $NaCu_5O_2(SeO_3)_2Cl_3$, and avdoninite $K_2Cu_5Cl_8(OH)_4 \cdot 2H_2O$. As was shown by the calculation results, these compounds could be good candidates in search and study of new quantum states.

The structure of the crystal sublattice of magnetic $Cu^{2+}$ ions in these minerals is composed of corner-sharing $Cu_4$ tetrahedra located on the kagome lattice. In each mineral, these layers have some specific peculiarities. The main difference consists in the fact that in averievite and ilinskite the $Cu_4$ tetrahedra are centered by oxygen ions, whereas in avdoninite they are void inside. The oxygen ions centering tetrahedra in averievite and ilinskite make the main contribution to antiferromagnetic components of spin interactions in $Cu_4$ tetrahedra. As we showed on the example of averievite, even slight changes in positions of the oxygen ions centering the tetrahedra could result to fluctuations of spin configurations. The absence of the inversion center in the crystal structure of averievite and the shift of tetrahedra-centering oxygen ions in the same direction indicate to the presence of a local electric polarization $[O_2Cu_5]^{6+}$ layers. Besides, averievite is characterized by a substantial anisotropy ($J3/J1 = 0.14$) of the strength of magnetic $J1$ and $J3$ couplings of the copper ions located above (Cu3) and below (Cu1) the kagome lattice with the Cu2 ions forming this lattice (Fig. 6a). In ilinskite, such an anisotropy is absent, as the Cu1 and Cu2 ions located above and below the kagome plane alternate. Structural phase transitions accompanied with magnetic transitions because of the effect of a lone electron pair of $Se^{4+}$ ions having a unilateral "umbrella" coordination in $SeO_3$-groups are possible in ilinskite.

As we have shown by comparison of averievite and ilinskite, at the removal of the oxygen ion from the tetrahedron center, the magnetic structure of the pyrochlore layer could transform into an openwork curled net woven from corner-sharing open AFM spin-frustrated tetrahedra ("butterflies"). The latter results in elimination of the kagome plane and disruption of the compliance of the magnetic structure with the crystal structure of the sublattice of magnetic $Cu^{2+}$ ions.

It is important to mention that two-dimensional frustrated antiferromagnetic spin-1/2 systems on the kagome lattice in three minerals we examined are formed by nonequivalent exchange interactions. We have revealed on the kagome lattice in averievite two ($J2$ and $J4$ ($J2/J4= 0.93$)) and in ilinskite four ($J6$, $J2$ ($J2/J6 = 0.57$), $J5$ ($J5/J6 = 0.37$), and $J8$ ($J8/J6 = 0.61$)) nonequivalent AFM exchange interactions, which do not differ significantly in strength in contrast to avdoninite. Three nonequivalent exchange interactions are present in avdoninite. Two of them ($J4_1$ and $J5_1$ ($J4_1/J5_1 = 0.57$)) are AFM and similar in strength, while the one ($J6_1$) is equal almost to 0, but could transform into the FM state.

The ratio of parameters of AFM interactions on the kagome lattice in averievite ($Cu_5O_2(VO_4)_2(CuCl)$) and ilinskite ($NaCu_5O_2(SeO_3)_2Cl$) is in compliance with the criteria of stabilization of the QSL state of the spin-1/2 Heisenberg models in case of isolated (separate kagome lattice) [58-65]. However, in averievite and ilinskite, the copper ions located above and below the lattice supplement the AFM spin-frustrated triangles of the kagome lattice until the AFM spin-frustrated tetrahedra. It remains unclear, which effect can be provided by these additional frustrated AFM couplings on the possibility of existence of the spin liquid in this case. We have not managed to identify theoretical criteria determining the possibility of existence of the spin liquid in quasi-two-dimensional and quasi-three-dimensional AFM frustrated magnetic structures.

It has been demonstrated that the structural-magnetic models of compounds built on the basis of calculations of magnetic couplings parameters by the crystal chemistry method can be effective in search of new functional magnetic materials.

**Acknowledgments**

The work was partially supported by the Program of Basic Research "Far East" (Far-Eastern Branch of the Russian Academy of Sciences), project no. 18–3–048.